\documentclass[11pt]{article}

\usepackage{geometry}  
\usepackage{placeins}
\usepackage[T1]{fontenc}  
\usepackage[utf8x]{inputenc}  
\usepackage{lineno}  
\usepackage{fancyhdr}  
\usepackage{enumitem}  

\usepackage[english]{babel}
\usepackage{csquotes} 
\usepackage[style=numeric-comp, maxbibnames=6, minbibnames=3, giveninits=true, doi=true, url=false, eprint=false, sorting=none]{biblatex}
\usepackage{hyperref}

\renewbibmacro{in:}{} 
\DeclareFieldFormat[article]{title}{#1} 
\DeclareFieldFormat[article]{journaltitle}{\textit{#1}} 
\DeclareFieldFormat[article]{volume}{\textbf{#1}} 
\DeclareFieldFormat[article]{number}{\mkbibparens{#1}} 
\DeclareFieldFormat[article]{pages}{#1} 

\DeclareNameAlias{author}{family-given} 

\renewcommand*{\cite}{\thinspace\supercite}

\usepackage{titling}  
\usepackage{mathtools}  
\usepackage{titlesec}  
\usepackage{lastpage}  
\usepackage{tabularx}
\usepackage{booktabs}

\usepackage{amsmath}  
\usepackage{wasysym}  
\usepackage{bbm}  
\usepackage{array}  
\usepackage{xr}  
\usepackage{verbatim} 
\usepackage{float} 
\usepackage{xcolor}
\usepackage{multirow}
\usepackage{amsmath,amssymb,amsfonts}
\usepackage{algorithm}
\usepackage{algpseudocode}
\usepackage{graphicx}
\usepackage{textcomp}

\usepackage[normalem]{ulem}  
\usepackage{xcolor}          
\usepackage{marginnote}      

\newif\ifshowchanges
\showchangesfalse  

\newcommand{\add}[2][]{\ifshowchanges \textcolor{blue}{#2} \marginnote{\textcolor{blue}{#1}} \else #2 \fi}

\DeclareMathOperator*{\argmin}{argmin\,}

\newcommand\nth[1]{$#1^{\mathrm{th}}$}

\definecolor{blue1}{RGB}{79, 113, 190}
\definecolor{orange1}{RGB}{239, 137, 51}
\definecolor{graynew}{gray}{0.88}
\definecolor{purple1}{RGB}{112, 48, 160}

\hypersetup{%
  pdfborderstyle={/S/U/W 1}
}

\titleformat{name=\section}{\normalfont\Large\bfseries}{}{0pt}{}
\titleformat{name=\subsection}{\normalfont\large\bfseries}{}{0pt}{}
\titleformat{name=\subsubsection}{\normalfont\normalsize\bfseries}{}{0pt}{}

\newcommand{\keywords}[1]{\textbf{Keywords}: {#1}}

\newcommand{\optincludegraphics}[2][]{}

\newcommand{\optinput}[1]{}

\newtagform{brackets}{[}{]}
\usetagform{brackets}

\newcommand{\thejournal}[1]{Magnetic Resonance in Medicine}

\title{\Large Robust multi-coil MRI reconstruction via self-supervised denoising}

\immediate\write18{texcount -sum=1,1,1,1,1,1,1 -merge -incbib -dir -utf8 \jobname.tex > \jobname.wcTotal }
\immediate\write18{texcount -sub=none -merge -incbib -dir -utf8 \jobname.tex | grep _main_ | grep -oE '[0-9]+' | python -c "import sys; print(sum(int(l) * w for l, w in zip(sys.stdin, [1, 1, 0, 0, 0, 0, 0])))" > \jobname.wcManuscript }
\immediate\write18{texcount -sub=none -merge -incbib -dir -utf8 \jobname.tex | grep Abstract | grep -oE '[0-9]+' | python -c "import sys; print(sum(int(l) * w for l, w in zip(sys.stdin, [1, 1, 0, 0, 0, 0, 0])))" > \jobname.wcAbstract }

\newcommand{\wcTotal}{\clearpage{\noindent\large{\bf Detailed Word Count} (not to be included for submission)}\verbatiminput{\jobname.wcTotal}}
\newcommand{\wcManuscript}{\input{\jobname.wcManuscript}}
\newcommand{\wcAbstract}{\input{\jobname.wcAbstract}}

\begin{document}

\begin{titlepage}
{\noindent\LARGE\bf \thetitle}

\bigskip

\begin{flushleft}\normalsize
	Asad Aali\textsuperscript{1,2},
	Marius Arvinte\textsuperscript{3},
	Sidharth Kumar\textsuperscript{2},
        Yamin I. Arefeen\textsuperscript{2,4},
        Jonathan I. Tamir\textsuperscript{2,5,6}

\end{flushleft}


\noindent
\begin{enumerate}[label=\textbf{\arabic*}]
\item {Department of Radiology\\Stanford University\\Stanford, CA, USA}
\item{Chandra Family Department of Electrical and Computer Engineering\\The University of Texas at Austin\\Austin, TX, USA}
\item {Intel Corporation\\Hillsboro, OR, USA}
\item {MD Anderson Cancer Center\\Houston, TX, USA}
\item {Dell Medical School Department of Diagnostic Medicine\\The University of Texas at Austin\\Austin, TX, USA}
\item {Oden Institute for Computational Engineering and Sciences\\The University of Texas at Austin\\Austin, TX, USA}
\end{enumerate}

\bigskip



\vfill


\end{titlepage}

\pagebreak

\begin{abstract}

\noindent\textbf{Purpose:} To examine the effect of incorporating self-supervised denoising as a pre-processing step for training deep learning (DL) based reconstruction methods on data corrupted by Gaussian noise. K-space data employed for training are typically multi-coil and inherently noisy. Although DL-based reconstruction methods trained on fully sampled data can enable high reconstruction quality, obtaining large, noise-free datasets is impractical.\\

\noindent\textbf{Methods:} We leverage Generalized Stein’s Unbiased Risk Estimate (GSURE) for denoising. We evaluate two DL-based reconstruction methods: Diffusion Probabilistic Models (DPMs) and Model-Based Deep Learning (MoDL). We evaluate the impact of denoising on the performance of these DL-based methods in solving accelerated multi-coil magnetic resonance imaging (MRI) reconstruction. The experiments were carried out on T2-weighted brain and fat-suppressed proton-density knee scans.\\

\noindent\textbf{Results:} We observed that self-supervised denoising enhances the quality and efficiency of MRI reconstructions across various scenarios. Specifically, employing denoised images rather than noisy counterparts when training DL networks results in lower normalized root mean squared error (NRMSE), higher structural similarity index measure (SSIM) and peak signal-to-noise ratio (PSNR) across different SNR levels, including 32dB, 22dB, and 12dB for T2-weighted brain data, and 24dB, 14dB, and 4dB for fat-suppressed knee data.\\

\noindent\textbf{Conclusion:} We showed that denoising is an essential pre-processing technique capable of improving the efficacy of DL-based MRI reconstruction methods under diverse conditions. By refining the quality of input data, denoising enables training more effective DL networks, potentially bypassing the need for noise-free reference MRI scans.\\

\noindent\keywords{MRI, accelerated reconstruction, self-supervised denoising, generative diffusion models, deep learning}

\end{abstract}

\newpage

\section{Introduction}
\label{sec:intro}

Magnetic resonance imaging (MRI) is a crucial imaging modality in clinical practice, offering detailed anatomical and functional information without ionizing radiation. Nonetheless, MRI is hindered by prolonged acquisition times, resulting in increased costs, patient discomfort, and motion-related artifacts that degrade image quality\cite{zaitsev2015motion}. To mitigate these issues, accelerated MRI techniques have been developed to reduce scan time by acquiring fewer data points\cite{deshmane2012parallel, lustig2007sparse, jaspan2015compressed, ye2019compressed, pal2022review, heckel2024deep}.

Clinically, the most prevalent method to accelerate MRI reconstruction is parallel imaging, which utilizes multiple receiver coils to acquire undersampled k-space data\cite{deshmane2012parallel, pruessmann1999sense, griswold2002generalized, ESPIRiT}. Although parallel imaging is effective, it suffers from noise amplification and residual aliasing at high acceleration rates, compromising image quality and diagnostic utility. Recent years have witnessed advances in applied supervised machine learning techniques for accelerated MRI reconstruction, broadly categorized into two primary approaches: end-to-end methods\cite{wang2016accelerating, hammernik2018learning, qin2018convolutional, aggarwal2018modl} and generative probabilistic methods\cite{jason_bayesianmri,tezcan_bayesianmri,jalal2021robust, chung2022score, luo2023bayesian, fan2024survey}. 

Competitive end-to-end deep learning (DL) methods train unrolled models by integrating a neural network architecture with conventional optimization to map undersampled k-space measurements to a high-quality coil-combined image. Although these methods typically deliver high-quality reconstructions, this performance is limited to in-distribution data (i.e., where the sampling pattern, field of view, and resolution are matched at training and at inference time), as the methods are conditioned on the measurement model\cite{fedorov2021tasting, cui2022self}. Generative methods\cite{jalal2021robust, chung2022score, song2020score, song2020improved, ho2020denoising, karras2022elucidating}, conversely, learn a prior distribution from fully sampled data and decouple prior learning from the forward model. Reconstruction is then formulated as a Bayesian inverse problem, for example as an approximation to posterior sampling. This approach allows greater flexibility, as the learned prior can be trained once and then applied to various inverse problems, regardless of the forward model. 

Both end-to-end and generative approaches rely on the availability of large, high-quality, fully sampled datasets, which can be difficult and costly to acquire\cite{coelho2013mr, haji2018validation}. Recent studies have introduced several self-supervised training and reconstruction methods that do not require fully sampled data\cite{senouf2019self, yaman2020self, yaman2020self2, millard2023theoretical,gan2023IEEETCI,yuyang2024spicer, daras2024ambient, aali2024ambient}. By creating specialized loss functions, these self-supervised techniques demonstrate that models trained on undersampled data can be utilized for high-quality reconstructions. In many practical scenarios involving low signal-to-noise ratio (SNR) data, such as low-field MRI, the available training data are inherently noisy\cite{koonjoo2021boosting, wang2024hidden, shimron2024accelerating}. Several self-supervised denoising methods have been introduced, circumventing the need for paired fully sampled noise-free images during training\cite{li1985stein, ramani2008monte, soltanayev2018training, metzler2018unsupervised, batson2019noise2self, krull2019noise2void, moran2020noisier2noise, kim2021noise2score, xiang2023ddm, elad2023image, mansour2023zero, pfaff2023self}. Self-supervised denoising methods for MRI applications have also been developed and applied, addressing measurement noise in MRI data and demonstrating high-quality MRI reconstructions\cite{tamir2019unsupervised, aggarwal2022ensure, millard2023clean, desai2023noise2recon, aali2023solving, aali2024gsure}. 

In this paper, we comprehensively explore the benefits of incorporating self-supervised denoising as a preprocessing step for DL-based MRI reconstruction methods. We first develop and implement a denoising pipeline using the Generalized Stein’s Unbiased Risk Estimate (GSURE)\cite{eldar2008generalized} and train denoiser networks for the fastMRI dataset\cite{zbontar2018fastmri}. Subsequently, in the context of generative models, we demonstrate that training on the denoised datasets produces qualitatively more accurate priors. Next, we assess the impact of self-supervised denoising on diffusion posterior sampling\cite{chung2023dps} and end-to-end reconstructions across various anatomies, training and inference SNR values, acceleration factors, and reconstruction times. Additionally, we investigate the qualitative performance of our proposed pipeline and show a reconstruction example with knee pathology across multiple SNR levels, demonstrating the impact of self-supervised denoising on diagnostic quality of reconstructions. This work is the first to simultaneously consider both end-to-end and generative models, providing a comprehensive view of how self-supervised denoising can enhance DL-based MRI reconstruction.


\section{Theory}
\label{sec:theory}

\subsection{Multi-Coil Magnetic Resonance Imaging}
\label{subsec:mc_mri_theory}

Multi-coil MRI is a linear inverse problem, with the forward operator $A$ comprising the coil sensitivity operator $S$, Fourier transform $F$, and sampling operator $P$\cite{pruessmann1999sense}. An image (without loss of generality assumed to be 2D) $X \in \mathbb{C}^{N_x \times N_y}$ is acquired using $N_c$ coils, with measurements in k-space. $N_x$ and $N_y$ are sample counts in frequency and phase encoding directions, respectively. The vectorized measurements are:
\begin{equation}
\label{eq:multicoil_accel_mri}
    y = PFSx + \eta,
\end{equation}
$x\in\mathbb{C}^{N_x N_y}$ is the vectorized image, $S \in \mathbb{C}^{N_c N_x N_y \times N_x N_y}$ denotes the sensitivity map operator that models the spatial sensitivities of the coils, $F$ is the Fourier transform applied across each coil, and $P \in \mathbb{C}^{(N_x N_y N_c / R) \times N_c N_x N_y}$ is the sampling operator that selects a subset of k-space locations, reducing the acquisition time by a factor of $R$.

The problem becomes ill-conditioned due to the undersampling of k-space, which results in missing information and noise amplification. Reconstructing the original signal $x$ from the undersampled k-space measurements $y$ is known as \textit{accelerated MRI reconstruction}. This problem is commonly formulated as an optimization problem, where the goal is to find the estimate $\tilde{x}$ that balances data consistency with regularization. One approach is to solve the regularized least squares problem:
\begin{equation}
\label{eq:regularized_LS_mri}
    \tilde{x} = \argmin_{x} \frac{1}{2} \underbrace{\|y - A x\|_2^2}_{\mathrm{data \ consistency}} + \lambda \underbrace{\mathcal{R}(x)}_{\mathrm{regularization}},
\end{equation}
where the first term, $\|y - A x\|_2^2$, enforces consistency between the measured data $y$ and the predicted measurements $A x$, assuming Gaussian noise. The second term, $\mathcal{R}(x)$, is a regularization term that incorporates prior knowledge about the distribution of $x$. Common choices include $\ell_1$-norm-based regularization, which promotes sparsity in certain transform domains, and total variation (TV) regularization, which encourages smoothness while preserving edges. Such regularizers are effective at reducing noise and undersampling artifacts while preserving key image features. 
This regularized formulation helps stabilize the solution, ensuring that the reconstruction is both accurate and robust to noise and undersampling artifacts. Deep learning methods utilize prior knowledge of the underlying distribution by learning from the training data.
When a statistical model for $p_X(x)$ is available, the regularizer can be expressed as the negative log-likelihood:
\begin{equation}
    \mathcal{R}(x) = -\log p_X(x).
\end{equation}

\subsection{Distribution Learning using Noisy Measurements}
\label{subsec:learning_priors_noisy_measurements}

In MRI, measurements are inherently noisy due to various physical and environmental factors, including the receiver coils' thermal noise, noise due to the body, and the acquisition system's imperfections. These noisy measurements introduce additional challenges when learning statistical priors about the ground-truth signals, as the noise corrupts the data, making it difficult to approximate the underlying signal distribution accurately. Equation~\eqref{eq:multicoil_accel_mri} describes a single set of measurements, such as those acquired during a single MRI scan. In practice, we often have access to a set of noisy but \textit{fully sampled} measurements obtained from different ground-truth signals $x^{(i)}$. \add[R1.C3]{The measurements $y^{(i)}$, including the signal and noise, are all complex-valued and are described as:}
\begin{equation}
\label{eq:measurement_dataset}
    y^{(i)} = F S^{(i)} x^{(i)} + \eta^{(i)},
\end{equation}
where $S^{(i)}$ denotes the coil sensitivity map corresponding to the $i$-th sample, $F$ is the 2D Fourier transform, and $\eta^{(i)}$ represents Gaussian noise, which is assumed to have zero mean and covariance matrix $C^{(i)}$. The superscripts indicate that each sample in the set can have different ground-truth signals, linear operators, and noise realizations.

We assume that each ground-truth signal $x^{(i)}$ is a realization of a continuous random variable $X$, which describes the distribution of signals across the population. For example, when working with 2D T2-weighted brain MRI scans, $X$ represents the distribution of all possible 2D T2-weighted brain scans in this setting. The log-probability density function of this random variable, $\log p_X(x)$, evaluates the likelihood of a particular realization $x$. The objective is to learn this prior distribution $\log p_X(x)$ and use it to inform the reconstruction process. Specifically, we integrate the prior into the regularized least squares framework for accelerated MRI reconstruction, as described in Equation~\eqref{eq:regularized_LS_mri}.

\subsection{Self-Supervised Denoising}
\label{subsec:gsure_theory}

One approach to learning a prior $\mathcal{R}(x)$ from noisy MRI measurements is applying \textit{self-supervised denoising}\cite{tamir2019unsupervised, aggarwal2022ensure, millard2023clean, desai2023noise2recon, aali2023solving, aali2024gsure} as a pre-processing step. The goal is to train a parametric model $g_\phi(A^\mathrm{H}y)$, parameterized by $\phi$, that maps a noisy estimate $A^\mathrm{H}y$ to its clean counterpart $x$, where $A^\mathrm{H}$ is the Hermitian transpose (adjoint) operator. The model $g_\phi$ is learned such that it minimizes the difference between the model's output and the clean data $x^{(i)}$. When ground-truth data $x^{(i)}$ are available for each noisy measurement $y^{(i)}$, the optimal model parameters $\phi^{*}$ could be learned by minimizing the supervised loss function based on empirical expectation:
\begin{equation}
    L_{\textrm{Sup}, \phi}(x) = \frac{1}{2} \sum_{i} \|g_\phi\left(A^\mathrm{H}y^{(i)}\right) - x^{(i)}\|_2^2.
\end{equation}
However, in most practical cases, ground-truth data are not available. Hence, we utilize GSURE\cite{eldar2008generalized} to obtain an unbiased estimate of the supervised loss based on the noisy measurements alone. \add[R1.C2, R2.C2]{We use GSURE instead of SURE\cite{tibshirani2015stein, metzler2018unsupervised} because the forward model $A$ is a linear operator and $A \neq I$.} This allows training without access to the true signal $x^{(i)}$. When the measurements $y^{(i)}$ are corrupted by independent and identically distributed (i.i.d.) Gaussian noise with a known covariance matrix $\sigma^2I$, the GSURE expression for the self-supervised loss is given by\cite{eldar2008generalized}:
\begin{equation}
\label{eq:gsure_loss}
    L_{\textrm{GSURE}, \phi}(A^\mathrm{H}y; \sigma) = \left|\left|g_\phi\left(\frac{A^\mathrm{H}y}{\sigma^2} \right)\right|\right|_2^2 + 2 \left[\mathrm{div}_{A^\mathrm{H}y}\left(g_\phi \left(\frac{A^\mathrm{H}y}{\sigma^2}\right)\right) - g_\phi^\mathrm{H}\left(\frac{A^\mathrm{H}y}{\sigma^2}\right) A^\dagger y\right].
\end{equation}
Here, $A^\dagger$ denotes the pseudo-inverse of the matrix $A$, and the term $\mathrm{div}_{x}(g_\phi(x))$ represents the divergence of the function $g_\phi$ at the point $x$, defined as:
\begin{equation}
\label{eq:div_def}
    \mathrm{div}_{x}(g_\phi(x)) = \sum_i \frac{\partial [g_\phi(x)]_i}{\partial x_i}.
\end{equation}

The computation of the full divergence in~\eqref{eq:div_def} requires evaluating the partial derivatives for each dimension of $x$, which is computationally expensive, especially for high-resolution MRI images. To address this, we approximate the divergence using a Monte Carlo method, following the approach in\cite{ramani2008monte}. The Monte Carlo estimate is computed using a single realization of zero-mean Gaussian noise $b \sim \mathcal{N}(0, I)$ and a small constant $\epsilon$. The approximation is given by:
\begin{equation}
\label{eq:div_approx}
    \mathrm{div}_x(g_\phi(x)) \approx b^\mathrm{T} \left(\frac{g_\phi(x + \epsilon b) - g_\phi(x)}{\epsilon}\right).
\end{equation}
This approximation drastically reduces the computational cost of evaluating the divergence while maintaining sufficient accuracy for training the denoising model. The GSURE loss function does not depend on the ground-truth data $x$, yet it has the remarkable property that its expectation is an unbiased estimate of the supervised loss. Specifically, it holds that for a fixed $A$:
\begin{equation}
    \mathbb{E}_{x, y} \left[ L_{\textrm{GSURE}, \phi}(A^\mathrm{H}y; \sigma) \right] = \mathbb{E}_{x, y} \left[ \left\|g_\phi\left(\frac{A^\mathrm{H}y}{\sigma^2}\right)\right\|_2^2 - 2 g_\phi^\mathrm{H}\left(\frac{A^\mathrm{H}y}{\sigma^2} \right) x \right],
\end{equation}
up to a constant factor\cite{eldar2008generalized}. In practice, the expectation $\mathbb{E}$ is taken over $x$, $y$, and $A$ according to\cite{tachella2022unsupervised}. The GSURE formulation serves as a reliable training loss for self-supervised learning in the absence of clean data, and it can be directly applied to scenarios where noisy measurements are available. GSURE loss can handle scenarios with varying noise levels $\sigma^{(i)}$ across different images, as is often the case in MRI acquisitions. In particular, this matches the setting where each coil can have different noise levels, in addition to different noise levels in each acquisition in a training set. This flexibility makes GSURE particularly well-suited for learning from noisy, real-world MRI data.

\begin{figure}
    \centering
    \includegraphics[scale=0.105]{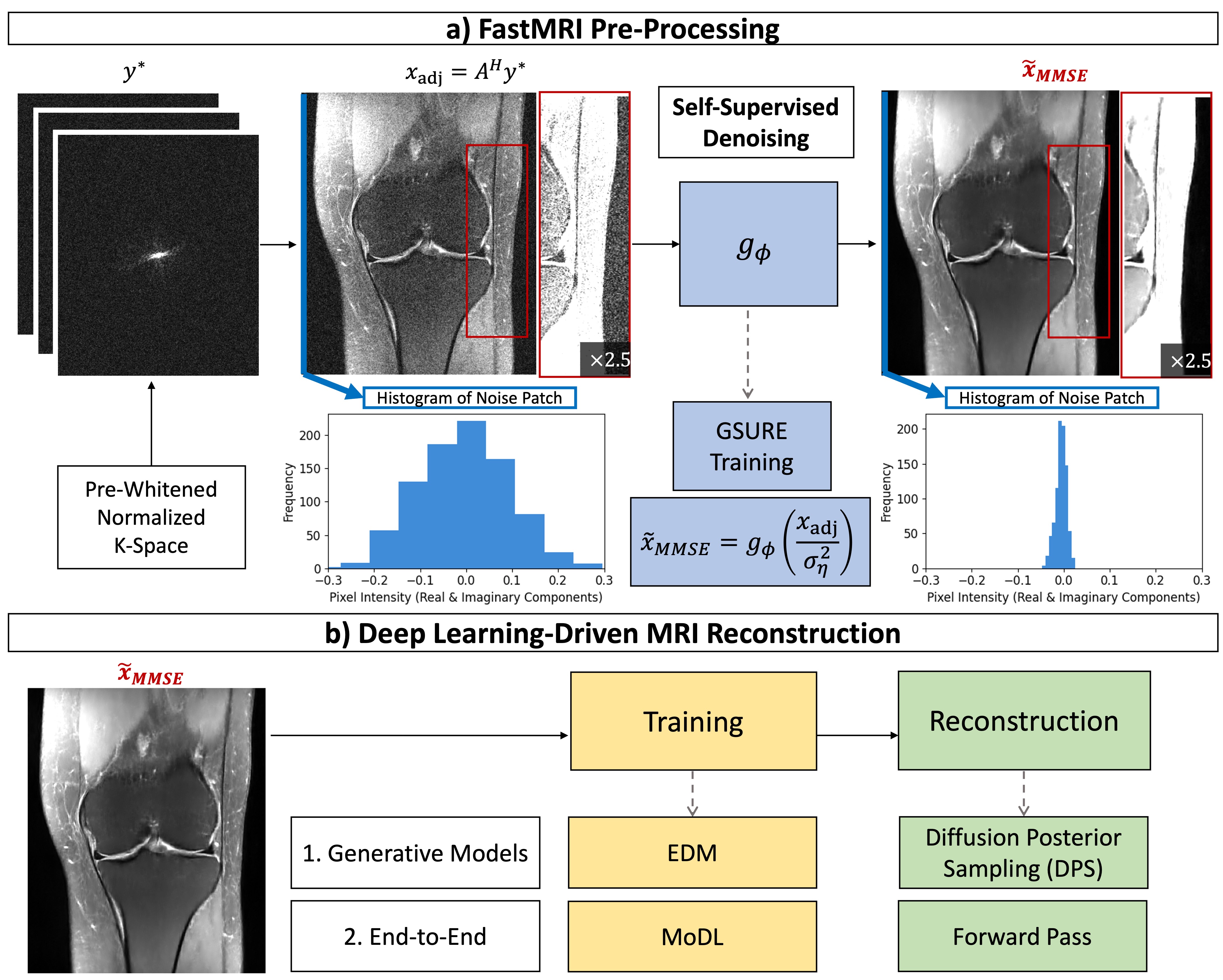}
    \caption{\scriptsize FastMRI Pre-Processing and Deep Learning-Driven MRI Reconstruction Pipeline. a) The pre-processing begins with pre-whitening and normalization of the raw k-space data. The whitened and normalized adjoint $A^\mathrm{H}y$ of the k-space is passed through the denoiser network $g_{\phi}$, outputting the MMSE denoised sample $\tilde{x}_\textrm{MMSE}$. \add[R1.C3]{We show the magnitude of a sample from the fastMRI dataset before and after denoising, including a histogram of the extracted noise patch to show the distribution of real and imaginary noise components.} b) The MMSE denoised data $\tilde{x}_\textrm{MMSE}$ are utilized for training deep learning networks using two methods: (1) Generative Models, and (2) End-to-End. Accelerated reconstruction is then performed utilizing: (1) Diffusion Posterior Sampling (DPS), and (2) MoDL Forward Pass.}
    \label{fig:pipeline}
\end{figure}

\section{Methods}
\label{sec:methods}

\subsection{Diffusion Probabilistic Models (DPMs) for Inverse Problems}
\label{subsec:diffusion_models_inverse}

Diffusion probabilistic models (DPMs) have recently shown potential as powerful generative models for solving ill-posed inverse problems, including accelerated MRI reconstruction. These models generate unconditional samples from a target distribution $p_0$ by iteratively refining noisy observations of the form $x_t = x_0 + \sigma(t) \eta_t$, where $\eta_t \sim \mathcal{N}(0, I)$ (for all $t$) represents Gaussian noise, and $\sigma(t)$ defines a time-dependent noise schedule. The primary objective of the DPM is to gradually denoise $x_t$ and generate an image $x_0 \sim p_0(x_0)$. This generative process is modeled by a stochastic differential equation (SDE)\cite{song2020score, karras2022elucidating}, which governs the backward denoising process:
\begin{gather}
    dx = -2\dot{\sigma}(t)(\mathbb{E}[x_0 | x_t] - x_t) dt + g(t) dw,
    \label{eq:backward_sde}
\end{gather}
where $w$ denotes the standard Wiener process, $g(t)$ is called the diffusion coefficient, and $\mathbb{E}[x_0 | x_t]$ is the model's estimate of the clean image at time $t$ based on the noisy input.

In inverse problem settings, the challenge is to recover $x_0$ from noisy, undersampled measurements $y$, which requires sampling from the posterior distribution $p(x_0 | y)$. This process combines prior knowledge of the distribution with the likelihood of the measurements given a clean image. The stochastic process for this is modeled by a corresponding backward stochastic differential equation (SDE), similar to the generative process:
\begin{gather}
    dx = -2\dot{\sigma}(t)(\mathbb{E}[x_0 | x_t, y] - x_t) dt + g(t) dw.
\end{gather}
Bayes' rule allows this posterior to be factored as $p(x_0 | y) \propto p(y | x_0) p(x_0)$, where $p(x_0)$ is the prior learned by the model, and $p(y | x_0)$ represents the likelihood of the measurements given $x_0$.

For most practical forward operators, computing the likelihood $p(y | x_t)$ in closed form is computationally intractable. Various approximation methods have been proposed to leverage DPMs for inverse problems\cite{jalal2021robust, song2021solving, kawar2022ddrm, chung2022improving, chung2023dps, feng2023score, diffusion_pnp}. An effective approximation technique is \textit{Diffusion Posterior Sampling (DPS)}\cite{chung2023dps}, which estimates $x_0$ using the noisy intermediate state $x_t$ and substitutes the intractable likelihood $p(y | x_t)$ with the conditional likelihood $p(y | \hat{x}_0)$, where $\hat{x}_0$ is an estimate of $x_0$ derived from the intermediate noisy state $x_t$. Specifically, DPS approximates $p(y | x_t)$ by $p(y | x_0=\mathbb{E}[x_0 | x_t])$, resulting in the following update equation:
\begin{gather}
    dx = -2\dot \sigma(t)\sigma(t)\left(\frac{\mathbb{E}[x_0 | x_t] - x_t}{\sigma(t)} + \gamma_t\underbrace{\nabla_{x_t}\log p(y | x_0=\mathbb{E}[x_0 | x_t])}_{\mathrm{likelihood \ term}}\right) dt + g(t) dw,
    \label{eq:dps_update}
\end{gather}
where $\gamma_t$ is a guidance parameter that modulates the influence of the data consistency term. The parameter $\gamma_t$ is typically adjusted based on the noise level $\sigma(t)$, with higher noise levels necessitating stronger guidance to ensure data fidelity.

DPMs allow for decoupling of prior learning from the measurement model, while their stochastic nature allows for uncertainty quantification by generating a distribution of possible solutions instead of just a single-point estimate. If a single-point estimate is desired, then these solutions can be averaged to form an approximate conditional expectation. This property is advantageous in scenarios with highly undersampled data, where multiple plausible reconstructions can exist.

\subsection{Model-Based Deep Learning (MoDL) for Inverse Problems}
\label{subsec:modl_inverse}

Model-based deep learning (MoDL)\cite{aggarwal2018modl} and other unrolled neural networks\cite{ongie2020deep} are end-to-end models that alternate between deep neural networks and differentiable optimization blocks. The MoDL framework in particular integrates the physics of the forward problem into a deep neural network architecture through the conjugate gradient algorithm. This makes it particularly well-suited for solving ill-posed inverse problems. The overall objective is to solve the following optimization problem (equation~\eqref{eq:regularized_LS_mri}):
\begin{gather}
    \tilde{x} = \argmin_{x} \underbrace{\|y - A x\|_2^2}_{\mathrm{data \ consistency}} + \lambda \underbrace{\|\mathcal{R}_\theta(x)\|^2}_{\mathrm{regularization}},
\end{gather}

where $A$ represents the forward operator, data consistency term $\|y - A x\|_2^2$ ensures the reconstruction matches the acquired measurements, regularization term $\|\mathcal{R}_\theta(x)\|^2$ imposes learned prior information to reduce artifacts, and $\lambda$ controls the trade-off between data fidelity and regularization. The regularization term $\mathcal{R}_\theta(x)$ is expressed as $x - \mathcal{D}_\theta(x)$, where $\mathcal{D}_\theta(x)$ is the estimate of $x$ after removal of noise and aliasing artifacts. The overall steps of this optimization are expressed as an alternating minimization between data consistency and denoising. The unrolled iterations comprising the full network are trained end-to-end, with the CNN-based denoiser $D_\theta$ shared across all iterations. $D_\theta$ is applied after the data-consistency update to remove residual artifacts, generating a refined reconstruction at each iteration, giving the objective:
\begin{gather}
    \tilde{x} = \argmin_{x} \|y - A x\|_2^2 + \lambda \ \|x - \mathcal{D}_\theta(x)\|^2,
\end{gather}

To train the supervised MoDL network, we utilize the normalized root mean squared error (NRMSE) objective between reconstructed images $\tilde{x}_i = H_\theta(y_i, A_i)$ and the fully sampled images $x_i$:
\begin{equation}
    L(x_i, \tilde{x}_i) = \frac{\| x_i - H_\theta(y_i, A_i) \|_2}{\| x_i \|_2},
\end{equation}
where $x_i$ are the fully sampled images, while $H_\theta(y_i, A_i)$ are the corresponding reconstructions generated by the MoDL network given the undersampled measurements $y_i$ and forward operator $A_i$. Here, $H_\theta$ refers to the entire MoDL network, which includes all unrolls of the data-consistency and denoising steps, while $\mathcal{D}_\theta$ specifically denotes the CNN-based denoiser applied in each iteration.

By incorporating the forward operator $A$ and the physics of the acquisition process directly into the architecture, MoDL ensures that the reconstructions are physically plausible and consistent with the measured data. Moreover, unrolling the iterative process allows the network to learn a reconstruction procedure that mimics traditional model-based methods, while CNN provides a flexible, data-driven regularization. However, a notable drawback of the MoDL framework is that the forward model $A$ is coupled during training, meaning that the model may not generalize well when the forward operator differs during testing or deployment, leading to generalization errors\cite{aggarwal2018modl}.

\subsection{Proposed Approach}
\label{subsec:proposed-approach}

To demonstrate proof-of-principal, we use the fastMRI\cite{zbontar2018fastmri} dataset as the source of our training data, which is preprocessed following the appropriate forward model defined in equation~\eqref{eq:multicoil_accel_mri}. \add[R2.C4]{Although fastMRI data are fully sampled, they are inherently noisy. Typically, a noise prescan is available that can be used for pre-whitening\cite{kellman2005image}. As fastMRI does not provide this information, we use noise-only patches from the multi-coil images to estimate the multi-coil noise covariance matrix.} To prepare the data for GSURE-based denoising, we apply noise pre-whitening using the Berkeley Advanced Reconstruction Toolbox (BART)\cite{uecker2015berkeley} and normalize the k-space data to account for variations in signal intensity across scans. This step reduces variability in signal levels, ensuring that noise and signal intensities are comparable across scans. Formally, we define the pre-whitened, normalized k-space as:
\begin{equation}
\label{eq:mri_whitened}
    y^{*} = FSx^{*} + \eta^{*},
\end{equation}
where $\eta^* \sim N(0, \sigma^{2}_{\eta^*} I)$ represents the normalized noise term with variance $\sigma^{2}_{\eta^*}$. The core of our approach involves learning the perturbed distribution of minimum mean square error (MMSE) denoised data. Although this formulation is distinct from directly learning the true distribution of clean data at arbitrary noise levels, we aim to decouple GSURE denoising from distribution learning. A schematic of the training pipeline is provided in Figure~\ref{fig:pipeline}. By accounting for noise, we hypothesize that it is possible to learn an approximate distribution of the clean images from denoised measurements. We propose a structured two-stage approach, trained sequentially:
\begin{enumerate}
    \item \textbf{MMSE Denoising}: We train a function $g_\phi$ to output minimum mean-square error (MMSE)-denoised, coil-combined images $\tilde{x}_{MMSE}$ from noisy multi-coil k-space data. This function is learned using the self-supervised GSURE training objective described in equation~\eqref{eq:gsure_loss}, where the GSURE network takes the adjoint of the measurements as input, scaled by the estimated noise variance.
    \item \textbf{Deep Learning-Driven MRI Reconstruction}: We train deep neural networks on the denoised outputs of $g_\phi$ to learn the distribution for subsequent accelerated MRI reconstruction:
        \begin{itemize}
            \item \textbf{DPM-based Reconstruction}: We utilize DPMs trained using the Elucidating Diffusion Models (EDM)\cite{karras2022elucidating} architecture and perform reconstruction using DPS\cite{chung2023dps}.
            \item \textbf{MoDL-based Reconstruction}: We use supervised MoDL\cite{aggarwal2018modl} (end-to-end).
        \end{itemize}
\end{enumerate}

We use terms \textbf{Naive-EDM} and \textbf{GSURE-EDM} to define EDM-based networks trained on noisy vs GSURE-denoised data, respectively. We then define two distinct DPM-based reconstruction methods: (1) \textbf{Naive-DPS} which leverages Naive-EDM priors for reconstruction using DPS, and (2) \textbf{GSURE-DPS} which leverages GSURE-EDM priors for reconstruction using DPS. Finally, we define terms for two distinct MoDL methods: (1) \textbf{Naive-MoDL} which leverages MoDL network trained on noisy data for one-step reconstruction, and (2) \textbf{GSURE-MoDL} which leverages MoDL network trained on GSURE-denoised data for one-step reconstruction. Each reconstruction method is evaluated by comparing the effectiveness of training on noisy versus GSURE-denoised data. 

\section{Experimental Setup}
\label{sec:experiments}

\subsection{Data} For training, we utilize k-space data from fully sampled T2-weighted brain and fat-suppressed proton-density knee scans from fastMRI\cite{zbontar2018fastmri}. We extract 5 central slices from each volume for both anatomies, resulting in a total of $10,000$ brain slices and $2,000$ knee slices. The k-space data are first converted into coil images by applying the inverse Fourier transform (IFFT), and these coil images are then cropped to matrix sizes of $384 \times 320$ for brain and $440 \times 368$ for knee.

After noise pre-whitening, we normalize k-space data by first cropping a $24\times24$ center region of k-space, then reconstructing this auto-calibration signal (ACS) region using the root sum-of-squares (RSS). We use the \nth{99} percentile of the ACS reconstruction to normalize pre-whitened k-space. We define the SNR of the pre-whitened and normalized dataset $y^*$ as:
\begin{equation}
    \text{SNR} = 10 \log_{10} \left(\frac{1}{\sigma^{2}_{\eta^{*}}}\right),
\end{equation}
where $\sigma_{\eta^*}^2$ represents the noise variance after pre-whitening and normalization. The brain dataset has an average native SNR of $32$ dB, while the knee dataset has an average SNR of $24$ dB.

For accelerated reconstruction, we retrospectively undersample 100 examples per anatomy using the same pre-processing steps. Random sampling is performed in the phase-encode direction with acceleration factors of $R=4$ and $R=8$, and a fully sampled $24 \times 24$ calibration region. We add additional noise to the k-space data to simulate different SNR levels. This is appropriate because the data were pre-whitened.

\subsection{Evaluation Metrics}
Reconstructions are compared with the native SNR fully sampled images and a mask is applied to the comparison images to only compare regions of the image containing anatomy. For quantitative assessment, we use (1) normalized root mean squared error (NRMSE) computed on the complex-valued images, (2) structural similarity index measure (SSIM), and (3) peak signal-to-noise ratio (PSNR). We emphasize that this evaluation is inherently biased, as native SNR images are inherently noisy\cite{wang2024hidden}; nonetheless, it provides a good proxy for reconstruction quality at lower SNR levels.

\subsection{Experiments}
We present a comprehensive evaluation of our proposed approach across different anatomies, noise levels, reconstruction techniques, and acceleration factors. We compare the reconstruction quality, denoising efficacy, and the impact of training on noisy versus GSURE-denoised data, providing quantitative and qualitative insights. We divide our experiments into three main categories:
\begin{enumerate}
    \item \textbf{Self-Supervised Denoising with GSURE}: We assess the effectiveness of GSURE denoising applied to pre-whitened and normalized noisy measurements. Independent denoisers were trained for both brain and knee datasets and for each SNR level: $32$ dB, $22$ dB, and $12$ dB for brain data, and $24$ dB, $14$ dB, and $4$ dB for knee data. After training, the denoisers were applied to the adjoint of the k-space data $x_{\textrm{adj}}$ to generate MMSE-denoised estimates $\tilde{x}_{MMSE}$.
    \item \textbf{Unconditional Sampling with EDM}: We examine the performance of EDM-based priors trained on noisy vs GSURE-denoised data by conducting qualitative evaluation of the generated scans across each anatomy and SNR level.
    \item \textbf{Accelerated MRI Reconstruction}: We evaluate the performance of our approach on undersampled and noisy k-space data using various reconstruction techniques. These experiments were designed to assess how well the models manage undersampling across different SNR levels. These experimentes are further divided for a comprehensive exploration of reconstruction quality and efficiency: (1) MRI reconstruction with DPS, (2) Knee pathology reconstruction with DPS, (3) MRI reconstruction speed with DPS, and (4) MRI reconstruction with MoDL. \add[R2.C5]{All DPS-based reconstructions were obtained by averaging five posterior samples, serving as an approximation to the MMSE estimate.} We also conducted statistical tests for experiments (1) and (4), to test whether there is a significant difference in the reconstruction performance of deep learning models trained on GSURE-denoised vs noisy data. We applied the non-parametric Wilcoxon signed-rank test for paired data, comparing reconstructions across varying scenarios including anatomies, acceleration factors, training and inference SNRs, giving us a total of $72$ paired tests. The tests were conducted with significance level $\alpha = 0.05$ and the Bonferroni corrections to adjust for multiple tests. The statistical analyses were performed using Python’s \textit{scipy.stats.wilcoxon}.
\end{enumerate}

\subsection{Implementation Details}
For GSURE training, we employed a UNet-style architecture\cite{ronneberger2015u} for both anatomies, comprising approximately $65$ million trainable parameters. The Adam optimizer was used for all experiments, with fixed learning rates ($\lambda$) tailored to the SNR levels of each dataset. Specifically, for brain data, we set $\lambda = 1 \times 10^{-4}$ across all SNR levels, while for knee data, $\lambda$ was set to $5 \times 10^{-7}$ for $24$ dB and $4$ dB SNR, and $5 \times 10^{-6}$ for the $14$ dB SNR level. The Monte Carlo divergence estimation $\epsilon$ was set to $0.001$ to balance computation accuracy and efficiency. All models were trained for $200$ iterations.

For DPM training, we followed the EDM loss formulation\cite{karras2022elucidating}. The noise distribution was defined using $\sigma_{min} = 0.002$ and $\sigma_{max} = 80$, with the learning rate dynamically adjusted according to the noise schedule and training step, as outlined in\cite{karras2022elucidating}. Each DPM model was trained over $3,000$ iterations, using the same UNet-style architecture as the GSURE training, with $65$ million parameters. Pre-processing and architectural configurations were consistent between GSURE and DPM experiments. Batch sizes were set according to GPU availability, with $15$ for brain experiments and $12$ for knee experiments on A100 GPUs, and $8$ for brain and $6$ for knee experiments on A40 GPUs. For DPS inference, a $500$-step linear noise schedule was employed, with $\sigma_{min} = 0.004$ and $\sigma_{max} = 10$, utilizing the Euler solver. The same network architecture as the GSURE and EDM models was used during inference, with hyperparameters fixed across all anatomies and SNR levels.

MoDL models were trained using a lighter UNet architecture containing approximately $1.9$ million trainable parameters. The number of unrolls was set to $6$ iterations per forward pass. Training was conducted over $10$ epochs with a batch size of $1$, using a learning rate of $\lambda = 0.0003$. The model’s input and output channels were designed to process the real and imaginary components of the MRI data, ensuring compatibility with the pre-processing steps in other experiments.

\clearpage
\begin{figure}
    \centering
    \includegraphics[scale=0.21]{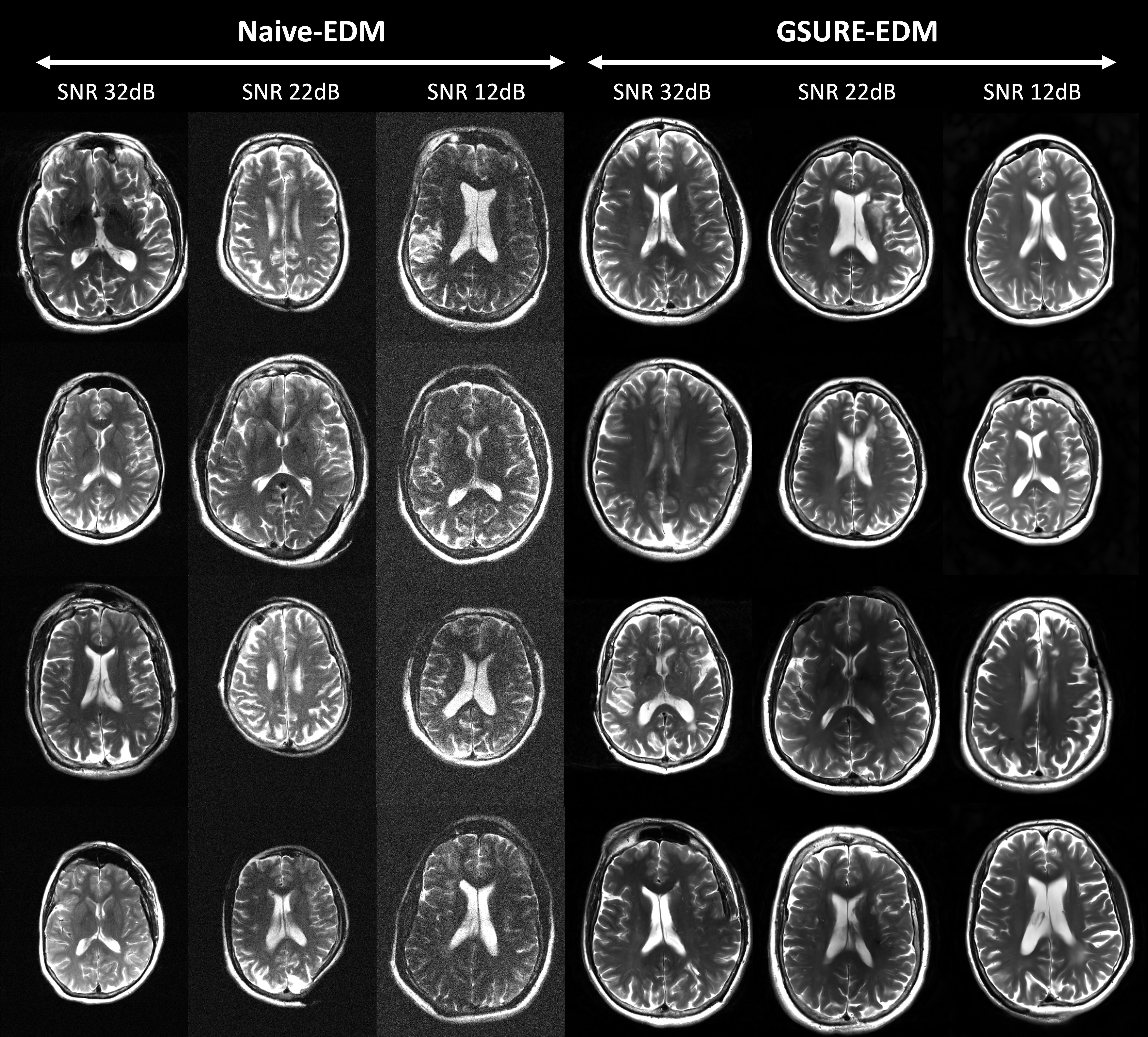}
    \caption{\scriptsize Unconditional T2-Weighted Brain images generated from EDM models trained on two datasets: a) Noisy (Naive-EDM), and b) GSURE denoised (GSURE-EDM). Across each column, we show prior samples across three different training SNR levels. Across each row, we show different realizations of images generated from the same distribution. We can observe that GSURE-EDM consistently generates qualitatively superior images, notably at lower SNR levels.
}
    \label{fig:brain_prior}
\end{figure}

\begin{figure}
    \centering
    \includegraphics[scale=0.21]{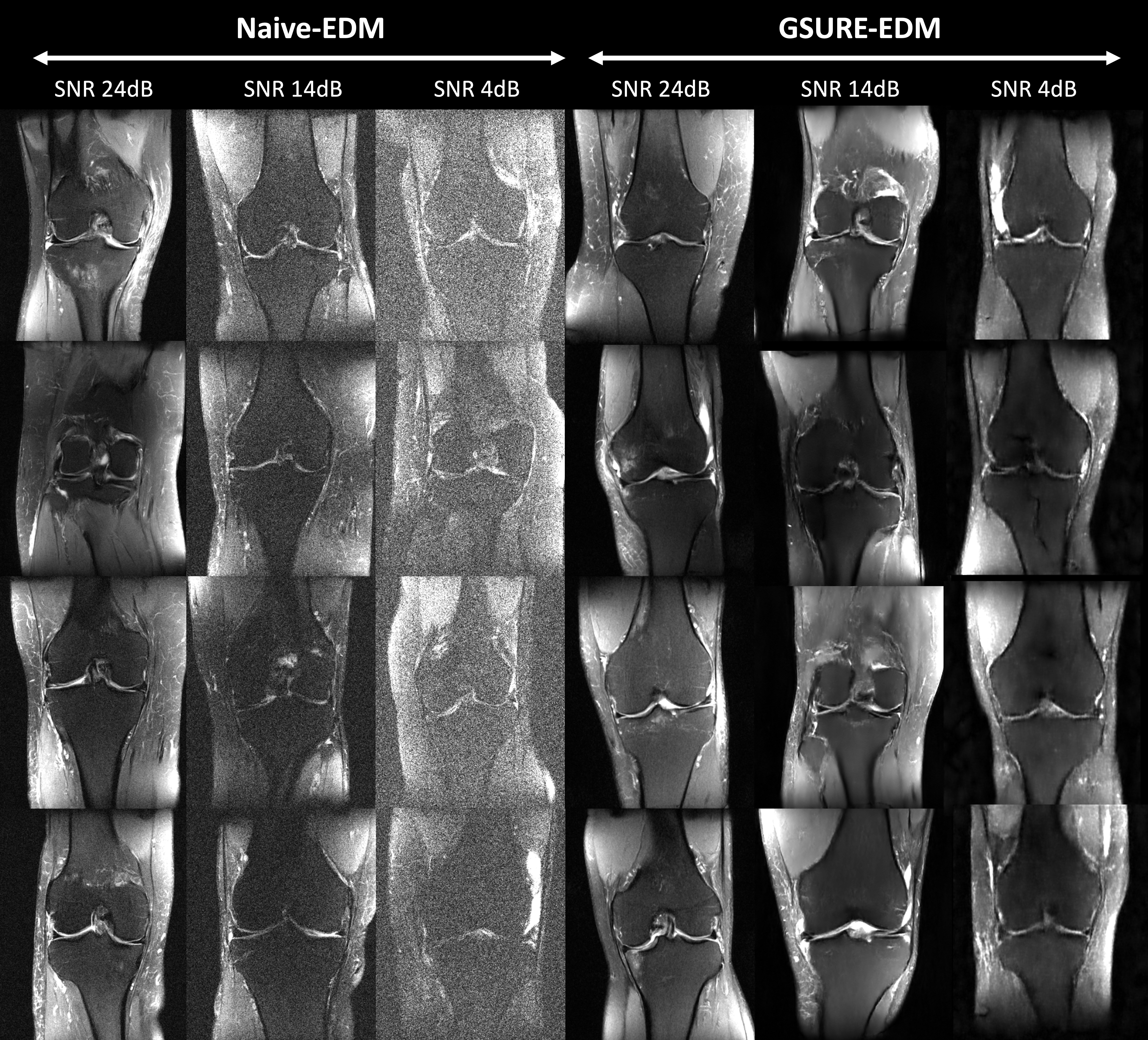}
    \caption{\scriptsize Unconditional Fat-Suppressed Knee images generated from EDM models trained on two datasets: a) Noisy (Naive-EDM), and b) GSURE denoised (GSURE-EDM). Across each column, we show prior samples across three different training SNR levels. Across each row, we show different realizations of images generated from the same distribution. We can observe that GSURE-EDM consistently generates qualitatively superior images, notably at lower SNR levels.
}
    \label{fig:knee_prior}
\end{figure}
\clearpage

\section{Results}
\subsection{Self-Supervised Denoising with GSURE}
\begin{figure}
    \centering
    \includegraphics[scale=0.29]{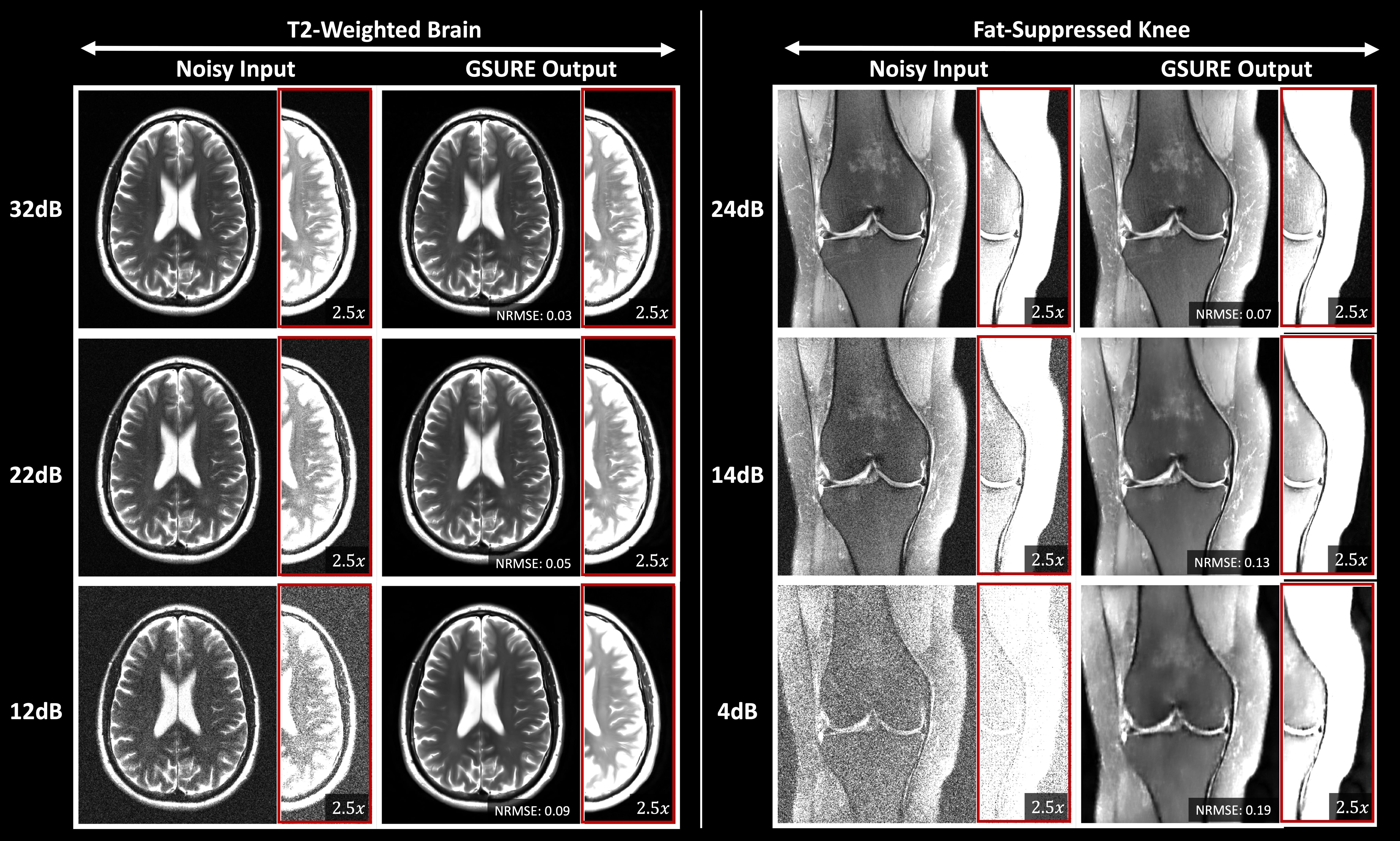}
    \renewcommand{\thefigure}{S1}
    \caption{\scriptsize Validation examples from the GSURE denoising experiment across T2-Weighted Brain and Fat-Suppressed Knee data at three SNR levels. Across each column under the appropriate anatomy, we show noisy input $A^\mathrm{H}y$ vs the output of the GSURE network $\tilde{x}_\textrm{MMSE} = g_{\phi}\left(\frac{A^\mathrm{H}y}{\sigma^2} \right)$. Across each row under the appropriate anatomy, we show the same validation example at different SNR levels, where row $1$ always shows data at the original native SNR (without additive noise). Overall, the experiment showcases the utility of GSURE denoising in improving the quality of training data, especially in low SNR cases.}
    \label{fig:denoising1}
\end{figure}

We report error metrics for denoising experiments averaged across $100$ validation examples in Supporting Table~\ref{tab:brain_knee_denoising_metrics}. Denoising performance degrades with decreasing SNR, with higher error values observed at lower SNRs. Supporting Figure~\ref{fig:denoising1} provides examples of the denoising experiments ($x_{\textrm{adj}}$ and $\tilde{x}_{MMSE}$) across varying SNRs. While noise reduction is evident, signal distortion becomes more pronounced as the SNR decreases, indicating a trade-off between noise removal and signal fidelity.

\subsection{Unconditional Sampling with EDM}
Figures~\ref{fig:brain_prior} and~\ref{fig:knee_prior} show prior samples from GSURE-EDM and Naive-EDM across each anatomy and SNR level. GSURE-EDM models consistently produce higher quality priors than Naive-EDM models. As SNR decreases, GSURE-EDM models offer qualitatively more accurate and realistic approximations of the fully sampled native SNR data, while Naive-EDM models retain noise.

\subsection{Accelerated MRI Reconstruction}

\begin{figure}
    \centering
    \includegraphics[scale=0.26]{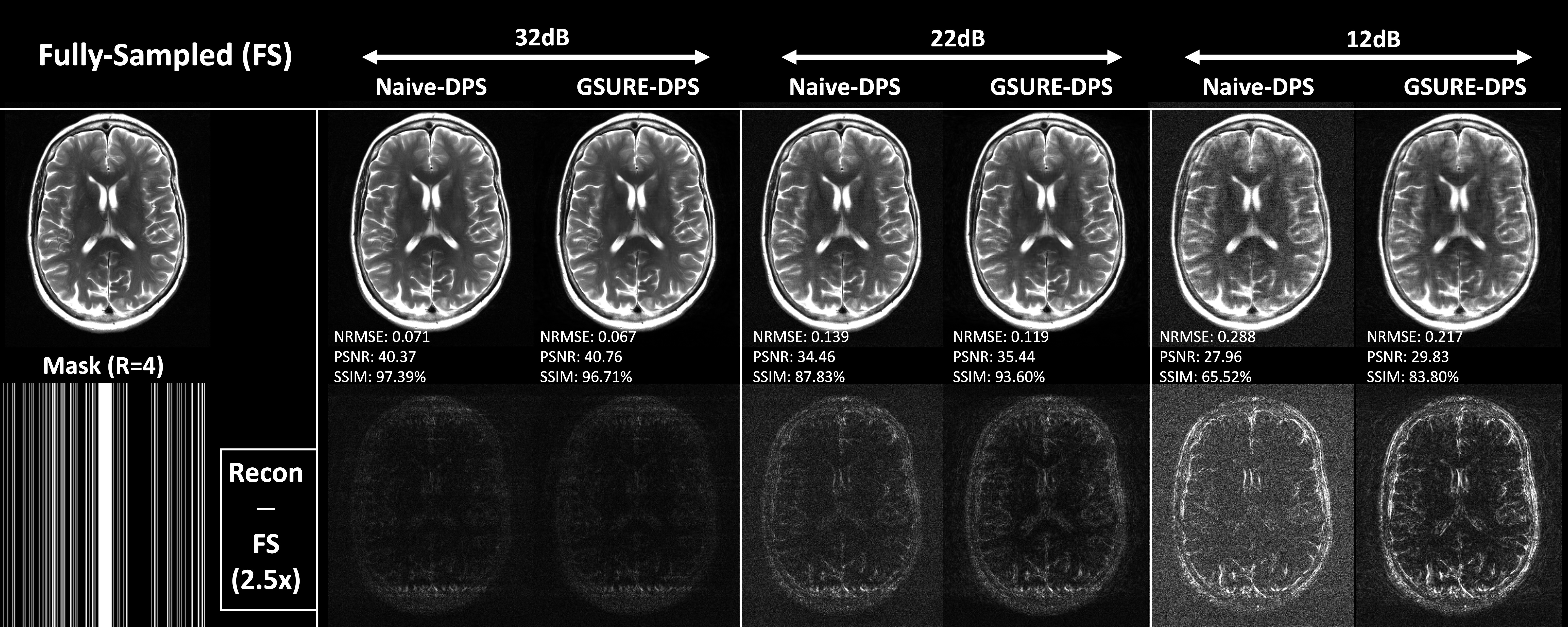}
    \caption{\scriptsize Conditional T2-Weighted Brain images with DPS as the reconstruction method, utilizing EDM models trained on two datasets: a) Noisy (Naive-DPS) and b) GSURE denoised (GSURE-DPS). Across columns, we show reconstructions across three training/inference SNR levels. In the first row, we show the reconstruction example with quantitative comparison metrics. In the second row, we show the difference between the reconstruction and fully sampled image at $2.5\times$ brightness. We can observe that GSURE-DPS consistently outperforms Naive-DPS, notably at lower SNR levels.}
    \label{fig:brain_posterior}
\end{figure}

\begin{figure}
    \centering
    \includegraphics[scale=0.095]{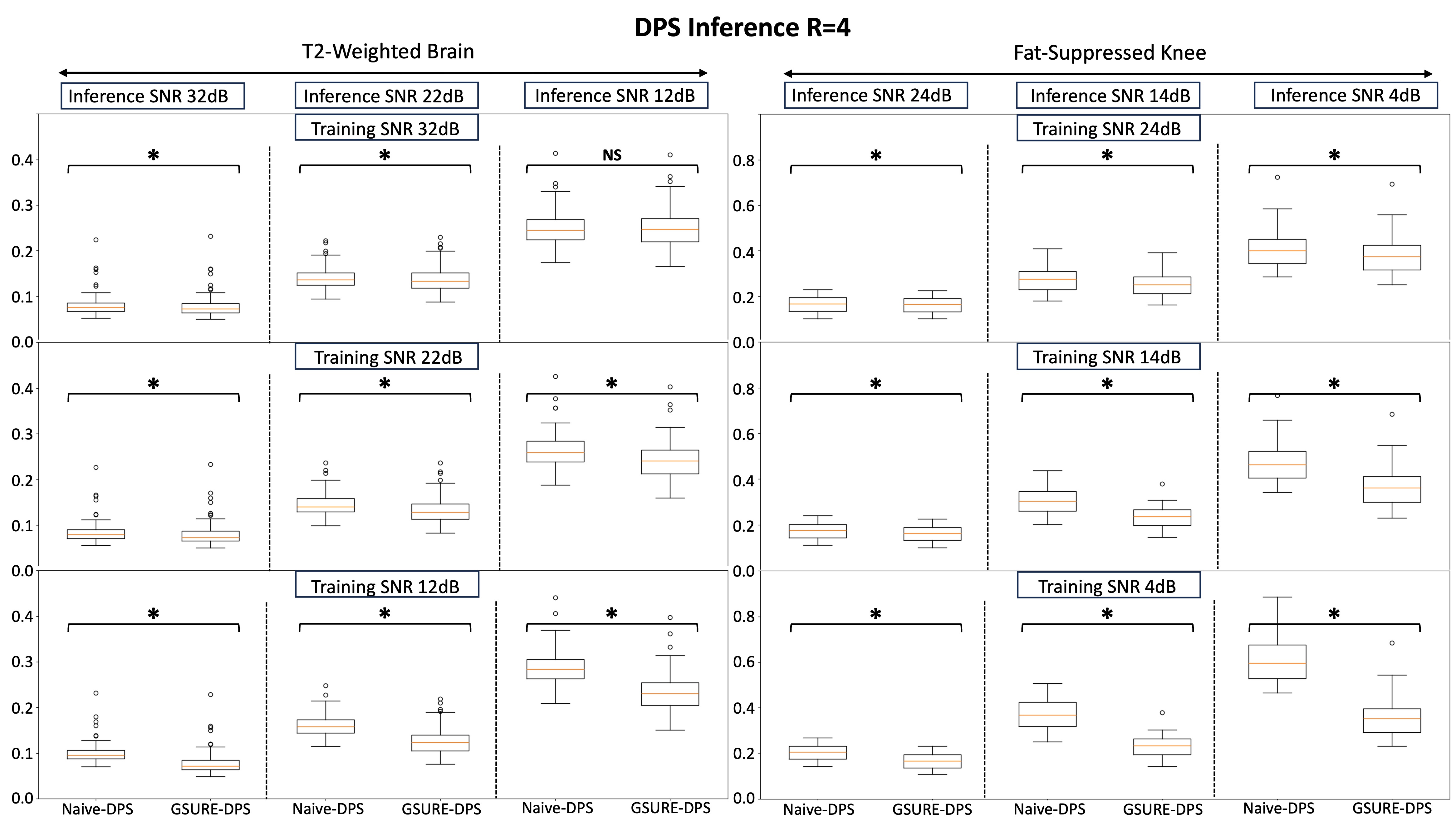}
    \renewcommand{\thefigure}{S2}
    \caption{\scriptsize Box plot of $100$ validation reconstruction examples (NRMSE) comparing Naive-DPS and GSURE-DPS at acceleration factor $R=4$, across: a) T2-Weighted Brain and Fat-Suppressed anatomies, b) training SNR levels, c) inference SNR levels. (\textbf{$^*$}statistically significant difference in reconstruction performance, NS: not significant).}
    \label{fig:dps_r4}
\end{figure}

\begin{figure}
    \centering
    \includegraphics[scale=0.095]{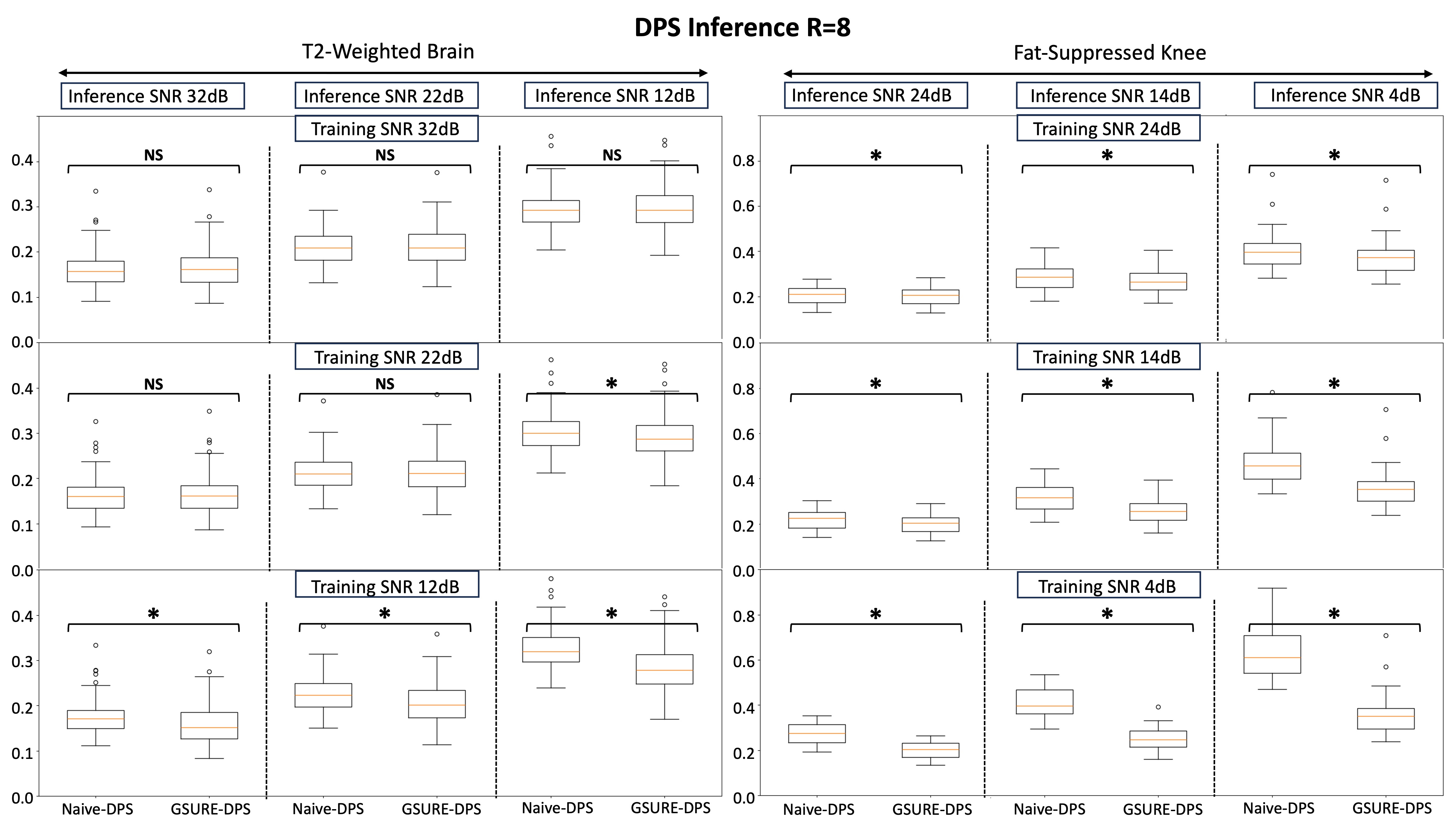}
    \renewcommand{\thefigure}{S3}
    \caption{\scriptsize Box plot of $100$ validation reconstruction examples (NRMSE) comparing Naive-DPS and GSURE-DPS at acceleration factor $R=8$, across: a) T2-Weighted Brain and Fat-Suppressed anatomies, b) training SNR levels, c) inference SNR levels. (\textbf{$^*$}statistically significant difference in reconstruction performance, NS: not significant).
}
    \label{fig:dps_r8}
\end{figure}

\begin{figure}
    \centering
    \includegraphics[scale=0.26]{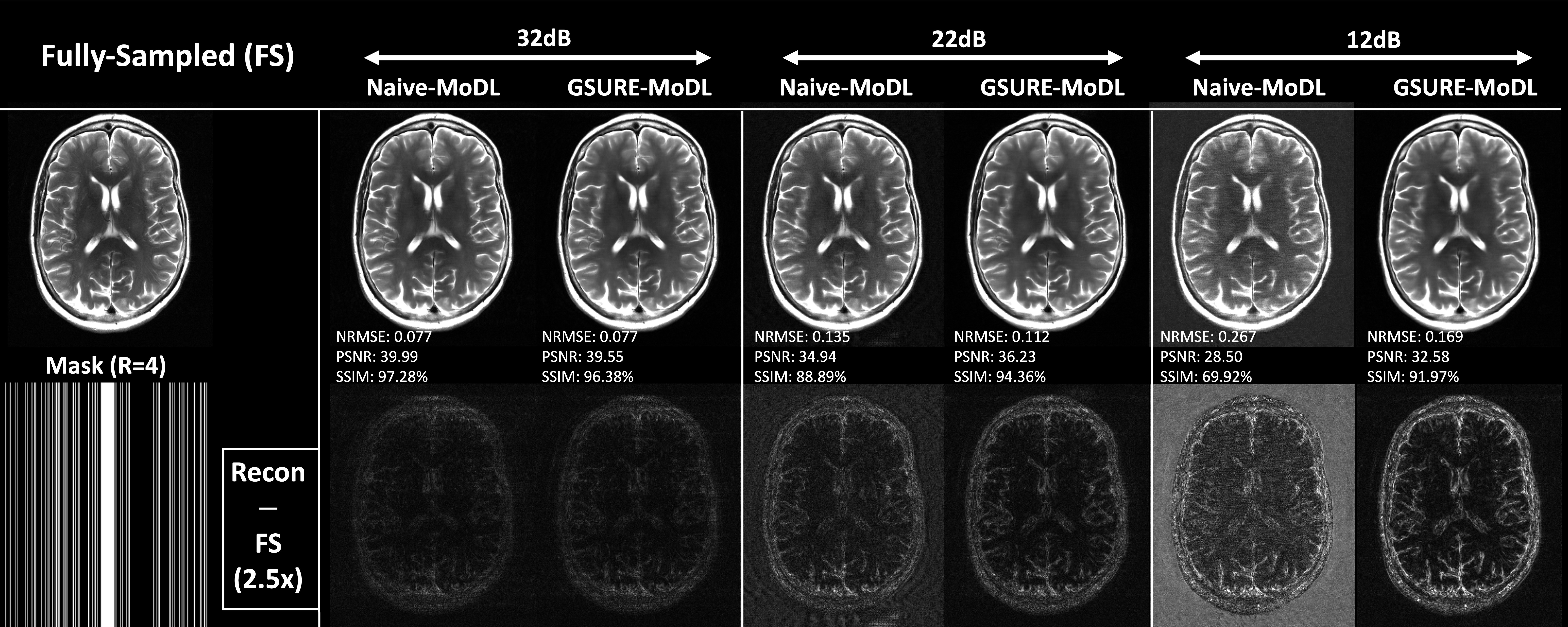}
    \renewcommand{\thefigure}{S7}
    \caption{\scriptsize \add[R1.C7]{T2-Weighted Brain reconstructions with MoDL, utilizing MoDL models trained on two datasets: a) Noisy (Naive-MoDL) and b) GSURE denoised (GSURE-MoDL). Across columns, we show reconstructions across three training/inference SNR levels. In the first row, we show the reconstruction example with quantitative comparison metrics. In the second row, we show the difference between the reconstruction and fully sampled image at $2.5\times$ brightness. We can observe that GSURE-MoDL outperforms Naive-DPS notably at lower SNR levels.}}
    \label{fig:brain_modl}
\end{figure}

\begin{figure}
    \centering
    \includegraphics[scale=0.26]{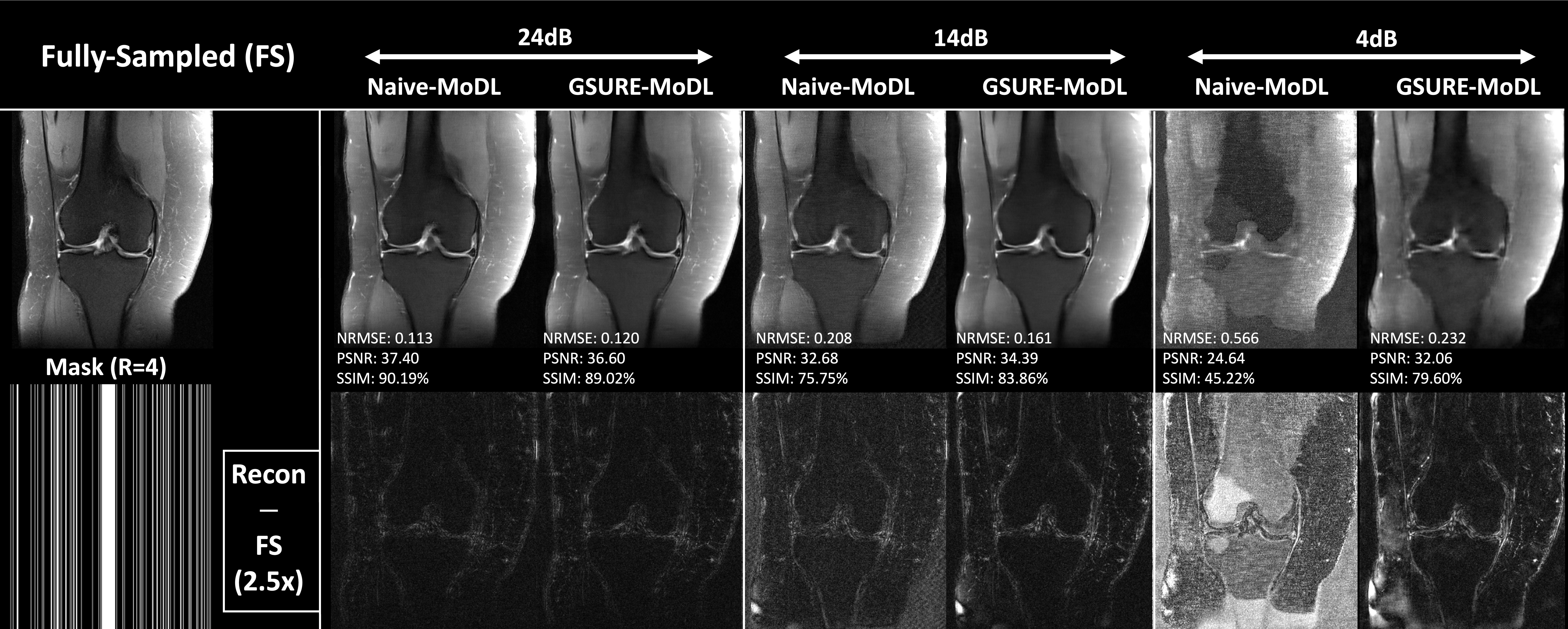}
    \renewcommand{\thefigure}{S8}
    \caption{\scriptsize \add[R1.C7]{Fat-Suppressed Knee reconstructions with MoDL, utilizing MoDL models trained on two datasets: a) Noisy (Naive-MoDL) and b) GSURE denoised (GSURE-MoDL). Across columns, we show reconstructions across three training/inference SNR levels. In the first row, we show the reconstruction example with quantitative comparison metrics. In the second row, we show the difference of the reconstruction and fully sampled image at $2.5\times$ brightness. We can observe that GSURE-MoDL outperforms Naive-DPS notably at lower SNR levels.}}
    \label{fig:knee_modl}
\end{figure}

Error metrics averaged across $100$ validation examples for each anatomy are reported under Tables~\ref{tab:brain_metrics} and~\ref{tab:knee_metrics}. Below, we describe the key experiments:
\begin{enumerate}     
    \item \textbf{MRI Reconstruction with DPS}:
    We compare the performance of Naive-DPS and GSURE-DPS and report error metrics across various inference conditions, averaging the results over five random seeds. Figures~\ref{fig:brain_posterior} and~\ref{fig:knee_posterior} display reconstructions and the difference images. At lower SNRs, GSURE-DPS consistently outperforms Naive-DPS both quantitatively and qualitatively. However, at the native SNR, both methods show comparable performance. The box plots in Figures~\ref{fig:dps_r4} and~\ref{fig:dps_r8} further illustrate the performance distributions across validation data, with GSURE-DPS consistently outperforming Naive-DPS at lower SNRs.
    
    \item \textbf{Knee Pathology Reconstruction with DPS}:
    We select a knee sample with visible pathology (meniscus tear) and average the results over five random seeds to ensure robustness and reduced variability. We evaluate inference SNR levels ranging from $24$ dB to $12$ dB, with models trained at a fixed SNR of $14$ dB. Figure~\ref{fig:knee_pathology_snr} shows reconstruction examples along with variance maps (across the five seeds) at $100\times$ brightness. GSURE-DPS exhibits stable reconstruction quality as inference SNR decreases, showing lower variation across seeds. GSURE-DPS outperforms Naive-DPS at lower inference SNRs as illustrated in Figure~\ref{fig:knee_pathology_deep}.

    \begin{figure}
    \centering
    \includegraphics[scale=0.26]{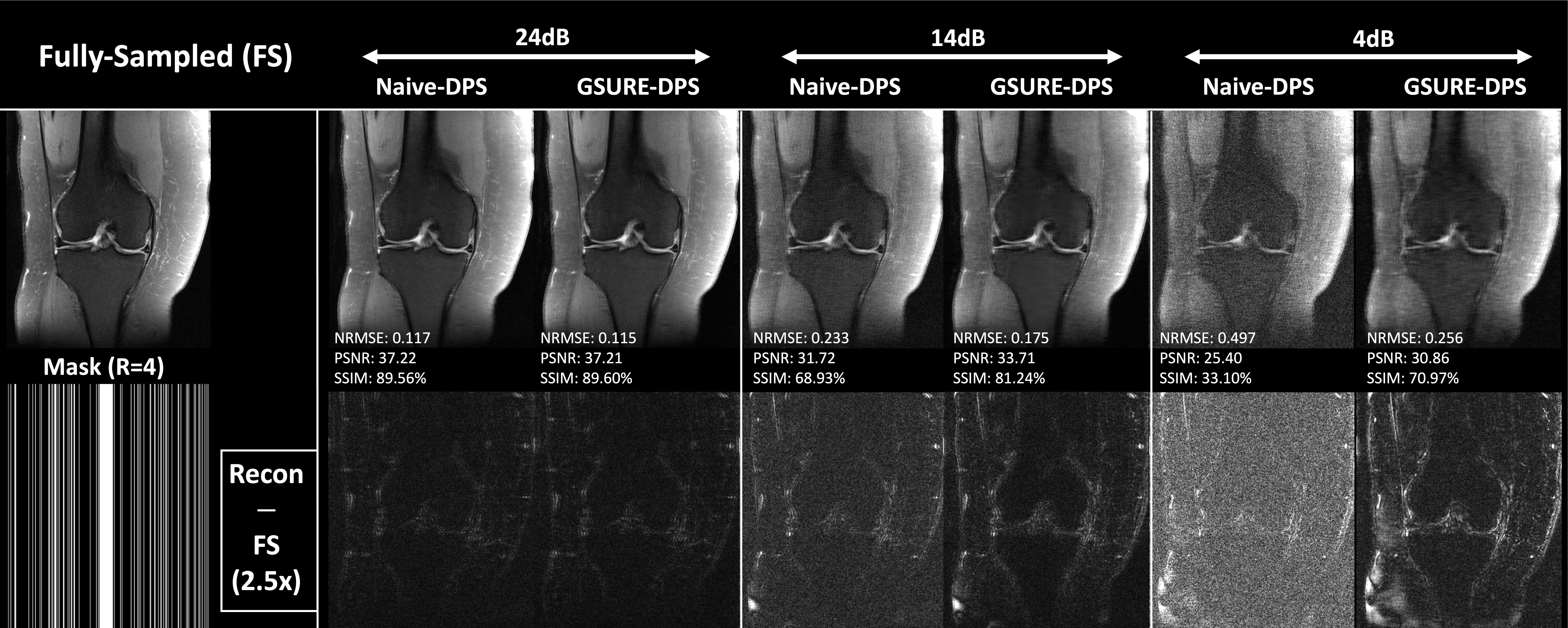}
    \caption{\scriptsize Conditional Fat-Suppressed Knee images with DPS as the reconstruction method, utilizing EDM models trained on two datasets: a) Noisy (Naive-DPS) and b) GSURE denoised (GSURE-DPS). Across columns, we show reconstructions across three training/inference SNR levels. In the first row, we show the reconstruction example with quantitative comparison metrics. In the second row, we show the difference of the reconstruction and fully sampled image at $2.5\times$ brightness. We can observe that GSURE-DPS consistently outperforms Naive-DPS, notably at lower SNR levels.}
    \label{fig:knee_posterior}
    \end{figure}

    \clearpage
    \begin{figure}
        \centering
        \includegraphics[scale=0.26]{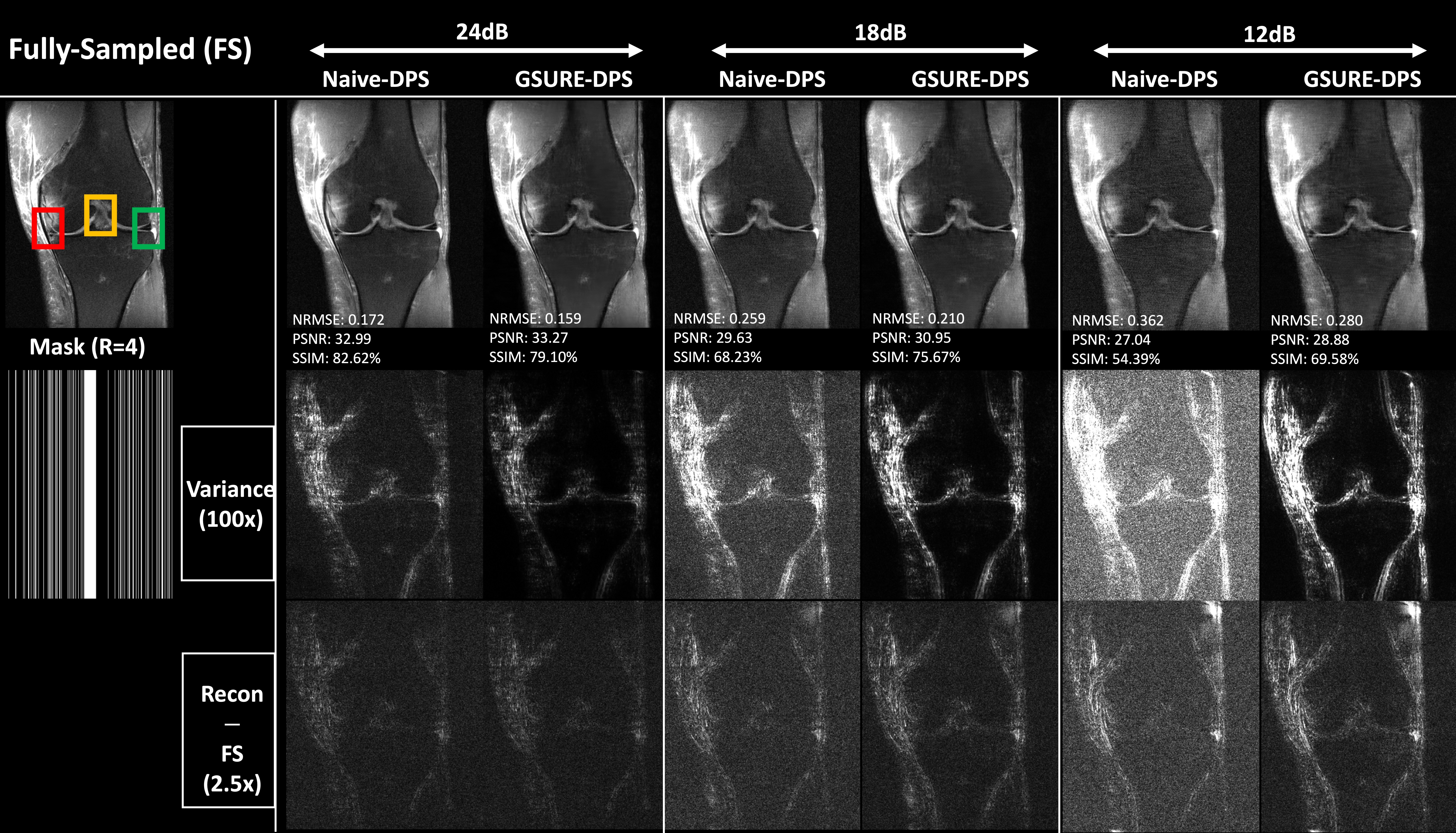}
        \caption{\scriptsize Conditional Fat-Suppressed Knee reconstruction with pathology (lateral meniscus tear) across multiple inference SNR levels, comparing Naive-DPS and GSURE-DPS trained at SNR $14$ dB. Across each column, we show reconstructions across inference SNR levels ranging from $24$ dB (native SNR) to $12$ dB. We show the relevant reconstruction example in the first row with quantitative comparison metrics. \add[R1.C6]{In the second and third rows, we show the variance of each reconstruction across five seeds at $100\times$ brightness, and the difference between the reconstruction and fully sampled image at $2.5\times$ brightness, respectively.} We can observe that GSURE-DPS reconstruction quality deteriorates more gracefully as inference SNR decreases.}
        \label{fig:knee_pathology_snr}
    \end{figure}
    
    \begin{figure}
        \centering
        \includegraphics[scale=0.26]{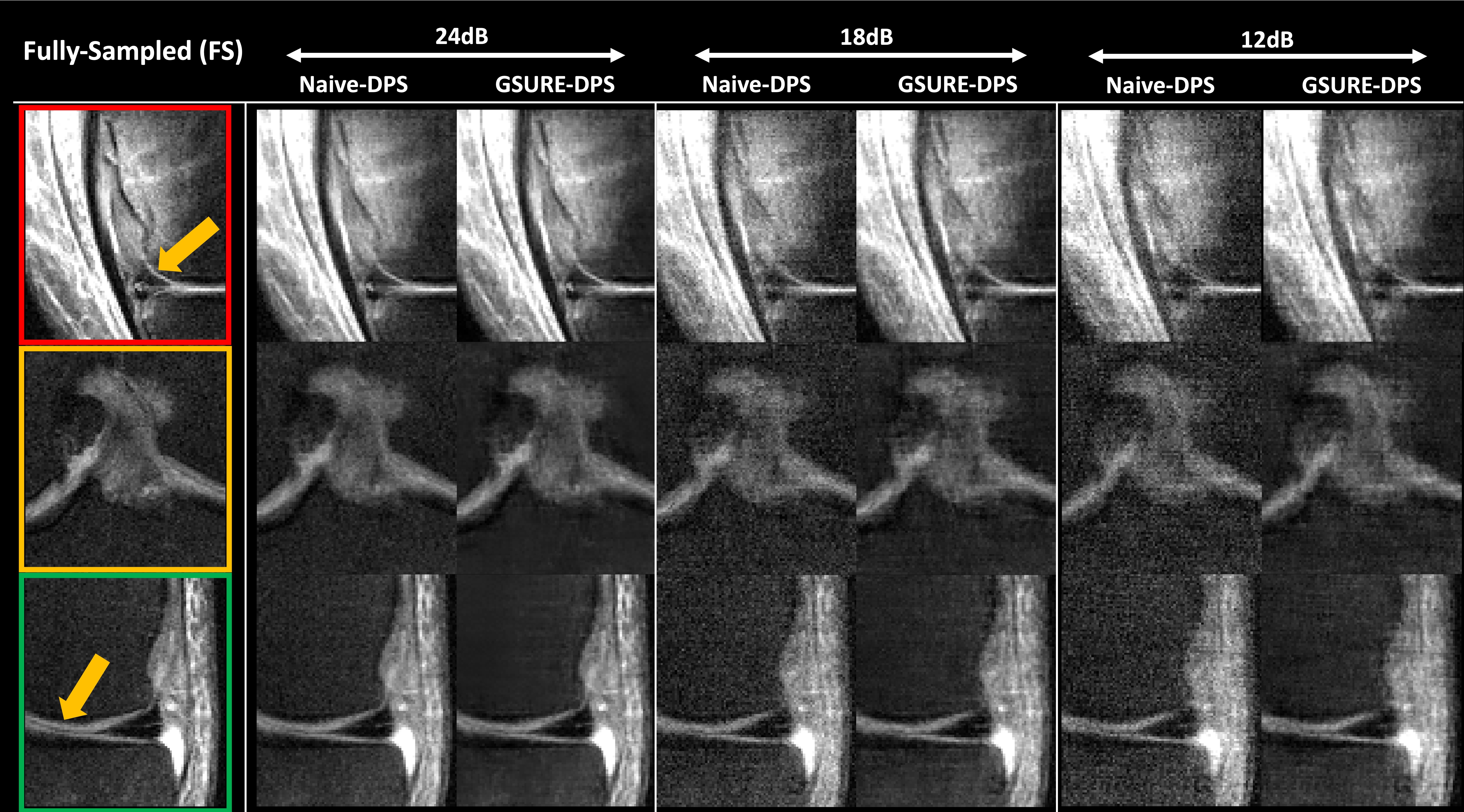}
        \caption{\scriptsize Zoomed-In conditional Fat-Suppressed Knee reconstruction with pathology (lateral meniscus tear) across multiple inference SNR levels, comparing Naive-DPS and GSURE-DPS trained at SNR $14$ dB. In each row, we zoom-in specific parts of the Knee. We can observe that GSURE-DPS reconstruction quality deteriorates more gracefully as inference SNR decreases.
    }
        \label{fig:knee_pathology_deep}
    \end{figure}
    \clearpage

    \begin{figure}
    \centering
    \includegraphics[scale=0.08]{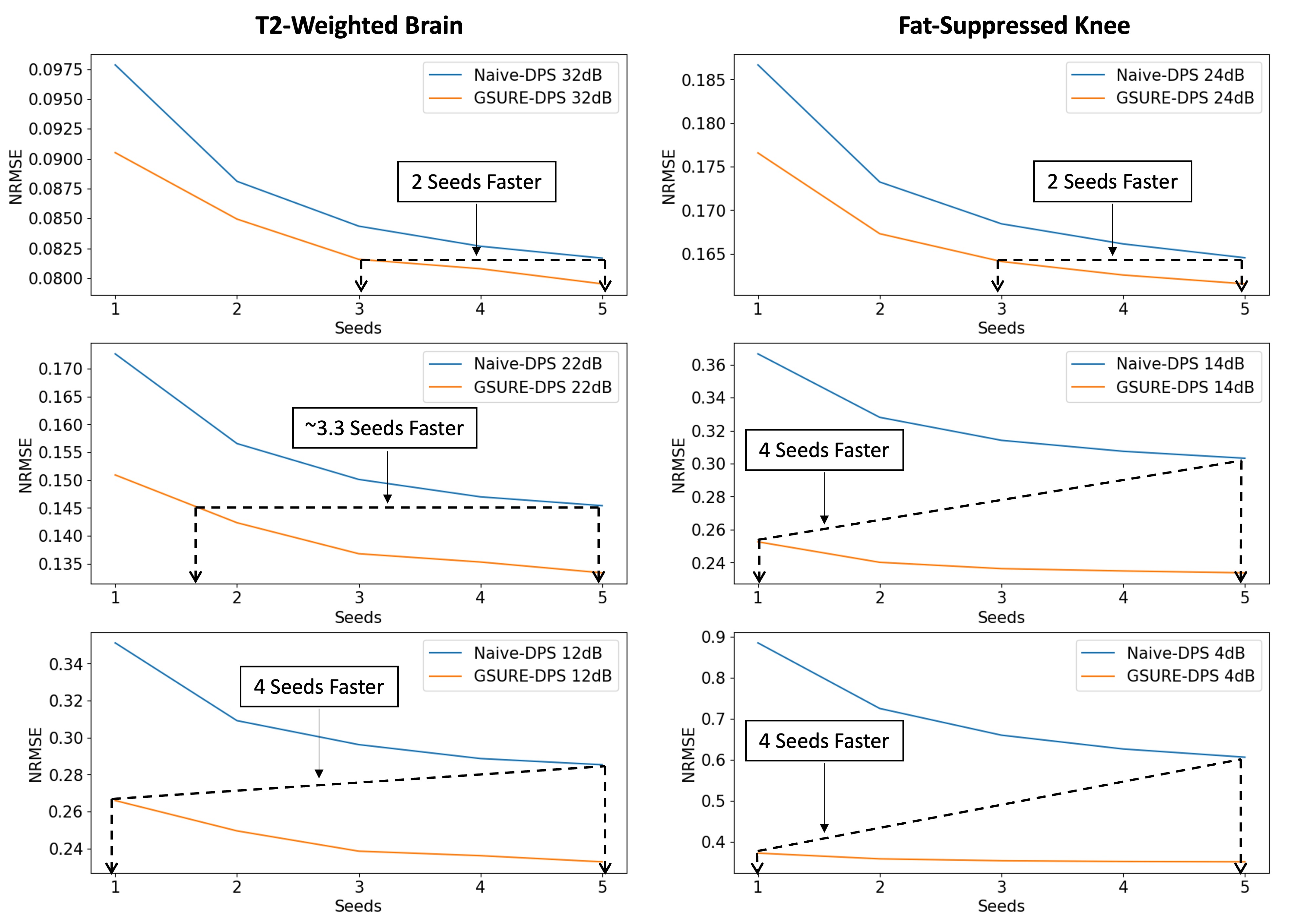}
    \caption{\scriptsize Reconstruction speed comparison between Naive-DPS and GSURE-DPS across T2-Weighted Brain and Fat-Suppressed Knee anatomies and three different SNR levels. Each plot compares Naive-DPS and GSURE-DPS, where each point on the line represents: a) on y-axis the average NRMSE across 100 validation examples, and b) on x-axis the number of seeds averaged to achieve that NRMSE. We can observe that GSURE-DPS reconstructions consistently require less averages to achieve a target NRMSE, showcasing that GSURE-DPS can improve reconstruction quality and enable faster reconstruction.}
    \label{fig:seed_inference}
    \end{figure}
    
    \item \textbf{MRI Training/Reconstruction Speed with DPS}:
    We evaluate reconstruction speed by measuring the number of posterior averages required to reach a target NRMSE. The error metrics were averaged across $100$ validation examples (Figure~\ref{fig:seed_inference}). GSURE-DPS consistently required fewer averages to achieve the target NRMSE. \add[R1.C4]{Furthermore, we also evaluate the impact of training set size on reconstruction performance in Figure~\ref{fig:training_speed}. We find that even when trained with only $10\%$ of samples from the original dataset, GSURE-DPS matches the reconstruction performance of Naive-DPS models, enabling more data-efficient training.}

    \item \textbf{MRI Reconstruction with MoDL}:
    \add[R1.C7, R2.C3]{We evaluate reconstructions across various SNRs and report error metrics averaged over $100$ examples. Figures~\ref{fig:brain_modl} and~\ref{fig:knee_modl} display reconstructions and the difference images. At lower SNRs, GSURE-MoDL consistently outperforms Naive-MoDL both quantitatively and qualitatively. Figure~\ref{fig:brain_ensure} and Table~\ref{tab:ensure_metrics} display that GSURE-MoDL outperforms ENSURE\cite{aggarwal2022ensure} when training SNR is low ($12$dB).} Figures~\ref{fig:modl_r4} and~\ref{fig:modl_r8} show the distributions of the NRMSE. GSURE-MoDL outperforms Naive-MoDL in low-SNR scenarios, while showing comparable performance at high inference SNRs. When training SNR is low and inference SNR is high (out-of-distribution), Naive-MoDL performs slightly better.
\end{enumerate}

\begin{table}[htbp]
\centering
\caption{\scriptsize All T2-Weighted Brain posterior reconstruction metrics. Across each inference SNR, acceleration factor, and reconstruction method, we \textbf{highlight} the training strategy with the best reconstruction performance (lowest error).}
\label{tab:brain_metrics}
\vspace{0.1in}
\resizebox{\textwidth}{!}{%
\begin{tabular}{ccc|ccc|ccc|ccc}
\toprule
\multirow{3}{*}{\shortstack{\textbf{Acceleration}\\\textbf{Factor}}} & \multirow{3}{*}{\shortstack{\textbf{Reconstruction}\\\textbf{Method}}} & \multirow{3}{*}{\shortstack{\textbf{Training}\\\textbf{SNR}}} & \multicolumn{9}{c}{\shortstack{\textbf{Inference SNR}}} \\
\cmidrule(lr){4-12}
 & & & \multicolumn{3}{c}{\textbf{32dB}} & \multicolumn{3}{c}{\textbf{22dB}} & \multicolumn{3}{c}{\textbf{12dB}} \\
\cmidrule(lr){4-6} \cmidrule(lr){7-9} \cmidrule(lr){10-12}
& & & \textbf{NRMSE $\downarrow$} & \textbf{SSIM $\uparrow$} & \textbf{PSNR $\uparrow$} & \textbf{NRMSE $\downarrow$} & \textbf{SSIM $\uparrow$} & \textbf{PSNR $\uparrow$} & \textbf{NRMSE $\downarrow$} & \textbf{SSIM $\uparrow$} & \textbf{PSNR $\uparrow$} \\
\midrule
\multirow{12}{*}{4} & \multirow{3}{*}{GSURE-DPS} & 32dB & 0.080 & 95.43 & 38.19 & 0.138 & 89.91 & 33.14 & 0.249 & 76.36 & 27.56 \\
 & & 22dB & 0.081 & 95.08 & 38.03 & 0.133 & 91.10 & 33.36 & 0.243 & 78.36 & 27.74 \\
 & & 12dB & \textbf{0.078} & 95.13 & \textbf{38.30} & \textbf{0.126} & \textbf{92.35} & \textbf{33.79} & \textbf{0.233} & \textbf{80.51} & \textbf{28.05} \\
 & \multirow{3}{*}{Naive-DPS} & 32dB & 0.082 & \textbf{96.04} & 37.99 & 0.140 & 88.55 & 33.07 & 0.248 & 73.57 & 27.70 \\
 & & 22dB & 0.085 & 95.59 & 37.70 & 0.145 & 86.63 & 32.81 & 0.262 & 69.36 & 27.31 \\
 & & 12dB & 0.101 & 93.35 & 36.12 & 0.161 & 83.37 & 31.96 & 0.285 & 64.88 & 26.68 \\
\cmidrule{2-12}
 & \multirow{3}{*}{GSURE-MoDL} & 32dB & 0.083 & 94.99 & 37.87 & 0.132 & 91.79 & 34.08 & 0.333 & 61.18 & 25.92 \\
 & & 22dB & 0.100 & 92.61 & 35.79 & \textbf{0.117} & \textbf{92.57} & \textbf{34.78} & 0.197 & 86.96 & 30.49 \\
 & & 12dB & 0.141 & 90.85 & 32.58 & 0.146 & 90.33 & 32.46 & \textbf{0.169} & \textbf{89.50} & \textbf{31.48} \\
 & \multirow{3}{*}{Naive-MoDL} & 32dB & \textbf{0.082} & \textbf{95.89} & \textbf{38.07} & 0.141 & 87.58 & 33.50 & 0.384 & 54.98 & 24.56 \\
 & & 22dB & 0.114 & 91.57 & 35.06 & 0.139 & 87.45 & 33.41 & 0.273 & 65.76 & 27.48 \\
 & & 12dB & 0.183 & 81.89 & 30.99 & 0.192 & 80.32 & 30.56 & 0.256 & 70.58 & 27.69 \\
\midrule
\multirow{12}{*}{8} & \multirow{3}{*}{GSURE-DPS} & 32dB & 0.164 & 89.30 & 30.95 & 0.214 & 83.94 & 28.52 & 0.297 & 73.04 & 25.59 \\
 & & 22dB & 0.165 & 89.22 & 30.92 & 0.213 & 84.59 & 28.54 & 0.293 & 74.76 & 25.67 \\
 & & 12dB & \textbf{0.159} & 89.82 & \textbf{31.22} & \textbf{0.206} & \textbf{86.00} & \textbf{28.80} & \textbf{0.286} & \textbf{76.97} & \textbf{25.85} \\
 & \multirow{3}{*}{Naive-DPS} & 32dB & 0.161 & \textbf{90.18} & 31.11 & 0.210 & 83.76 & 28.70 & 0.295 & 70.81 & 25.71 \\
 & & 22dB & 0.162 & 89.90 & 31.05 & 0.213 & 82.36 & 28.62 & 0.303 & 67.42 & 25.55 \\
 & & 12dB & 0.176 & 86.99 & 30.37 & 0.226 & 78.37 & 28.18 & 0.325 & 61.95 & 25.08 \\
\cmidrule{2-12}
 & \multirow{3}{*}{GSURE-MoDL} & 32dB & \textbf{0.132} & 92.27 & \textbf{33.19} & 0.170 & 88.40 & 31.22 & 0.348 & 58.66 & 25.18 \\
 & & 22dB & 0.149 & 89.64 & 31.76 & \textbf{0.160} & \textbf{90.20} & \textbf{31.37} & 0.227 & 80.42 & 28.80 \\
 & & 12dB & 0.183 & 86.19 & 29.84 & 0.186 & 86.47 & 29.79 & \textbf{0.204} & \textbf{86.76} & \textbf{29.29} \\
 & \multirow{3}{*}{Naive-MoDL} & 32dB & 0.133 & \textbf{92.44} & 33.12 & 0.177 & 83.25 & 30.86 & 0.368 & 55.36 & 24.65 \\
 & & 22dB & 0.151 & 88.81 & 31.96 & 0.166 & 85.93 & 31.30 & 0.258 & 69.38 & 27.68 \\
 & & 12dB & 0.211 & 78.82 & 29.03 & 0.215 & 78.04 & 28.94 & 0.265 & 70.83 & 27.11 \\
\midrule
\end{tabular}%
}
\end{table}

\begin{table}[htbp]
\centering
\caption{\scriptsize All Fat-Suppressed Knee posterior reconstruction metrics. Across each inference SNR, acceleration factor, and reconstruction method, we \textbf{highlight} the training strategy with the best reconstruction performance (lowest error).}
\label{tab:knee_metrics}
\vspace{0.1in}
\resizebox{\textwidth}{!}{%
\begin{tabular}{ccc|ccc|ccc|ccc}
\toprule
\multirow{3}{*}{\shortstack{\textbf{Acceleration}\\\textbf{Factor}}} & \multirow{3}{*}{\shortstack{\textbf{Reconstruction}\\\textbf{Method}}} & \multirow{3}{*}{\shortstack{\textbf{Training}\\\textbf{SNR}}} & \multicolumn{9}{c}{\shortstack{\textbf{Inference SNR}}} \\
\cmidrule(lr){4-12}
 & & & \multicolumn{3}{c}{\textbf{24dB}} & \multicolumn{3}{c}{\textbf{14dB}} & \multicolumn{3}{c}{\textbf{4dB}} \\
\cmidrule(lr){4-6} \cmidrule(lr){7-9} \cmidrule(lr){10-12}
& & & \textbf{NRMSE $\downarrow$} & \textbf{SSIM $\uparrow$} & \textbf{PSNR $\uparrow$} & \textbf{NRMSE $\downarrow$} & \textbf{SSIM $\uparrow$} & \textbf{PSNR $\uparrow$} & \textbf{NRMSE $\downarrow$} & \textbf{SSIM $\uparrow$} & \textbf{PSNR $\uparrow$} \\
\midrule
\multirow{12}{*}{4} & \multirow{3}{*}{GSURE-DPS} & 24dB & 0.162 & 83.72 & \textbf{34.50} & 0.248 & 72.25 & 30.91 & 0.375 & 59.44 & 27.07 \\
 & & 14dB & \textbf{0.160} & 81.06 & 34.39 & 0.234 & 74.00 & 31.22 & 0.358 & 62.96 & 27.33 \\
 & & 4dB & 0.166 & 80.21 & 33.97 & \textbf{0.231} & \textbf{74.50} & \textbf{31.33} & \textbf{0.351} & \textbf{63.98} & \textbf{27.70} \\
 & \multirow{3}{*}{Naive-DPS} & 24dB & 0.165 & \textbf{84.16} & 34.45 & 0.269 & 68.55 & 30.32 & 0.402 & 53.14 & 26.64 \\
 & & 14dB & 0.172 & 83.20 & 34.09 & 0.303 & 62.97 & 29.40 & 0.467 & 44.37 & 25.57 \\
 & & 4dB & 0.204 & 78.23 & 32.65 & 0.370 & 53.75 & 27.74 & 0.606 & 33.16 & 23.54 \\
\cmidrule{2-12}
 & \multirow{3}{*}{GSURE-MoDL} & 24dB & 0.160 & 82.36 & 34.37 & 0.269 & 69.56 & 30.71 & 0.824 & 29.52 & 20.96 \\
 & & 14dB & 0.185 & 75.22 & 32.88 & \textbf{0.215} & \textbf{75.09} & \textbf{32.15} & 0.397 & 60.65 & 27.83 \\
 & & 4dB & 0.415 & 70.96 & 24.77 & 0.255 & 71.58 & 30.11 & \textbf{0.315} & \textbf{70.49} & \textbf{29.71} \\
 & \multirow{3}{*}{Naive-MoDL} & 24dB & \textbf{0.159} & \textbf{84.59} & \textbf{34.82} & 0.341 & 59.34 & 28.69 & 1.202 & 21.46 & 17.40 \\
 & & 14dB & 0.180 & 82.92 & 33.91 & 0.296 & 66.44 & 29.79 & 0.804 & 30.18 & 20.51 \\
 & & 4dB & 0.303 & 72.12 & 29.29 & 0.375 & 63.17 & 27.46 & 0.890 & 35.54 & 19.55 \\
\midrule
\multirow{12}{*}{8} & \multirow{3}{*}{GSURE-DPS} & 24dB & 0.203 & 77.51 & \textbf{31.96} & 0.266 & 69.83 & 29.95 & 0.368 & 59.00 & 27.49 \\
 & & 14dB & \textbf{0.200} & 75.73 & 31.95 & 0.254 & 70.96 & 30.13 & 0.351 & 62.64 & 27.76 \\
 & & 4dB & 0.202 & 75.13 & 31.82 & \textbf{0.249} & \textbf{71.76} & \textbf{30.32} & \textbf{0.348} & \textbf{63.57} & \textbf{28.03} \\
 & \multirow{3}{*}{Naive-DPS} & 24dB & 0.207 & \textbf{77.65} & 31.88 & 0.282 & 66.84 & 29.55 & 0.395 & 53.21 & 26.95 \\
 & & 14dB & 0.220 & 75.81 & 31.48 & 0.315 & 60.97 & 28.76 & 0.458 & 44.45 & 25.82 \\
 & & 4dB & 0.275 & 66.09 & 29.81 & 0.413 & 47.35 & 26.60 & 0.627 & 31.25 & 23.21 \\
\cmidrule{2-12}
 & \multirow{3}{*}{GSURE-MoDL} & 24dB & \textbf{0.195} & 77.44 & 32.33 & 0.279 & 68.01 & 30.01 & 0.795 & 30.46 & 21.30 \\
 & & 14dB & 0.211 & 72.41 & 31.40 & \textbf{0.236} & \textbf{72.52} & \textbf{30.91} & 0.398 & 59.73 & 27.54 \\
 & & 4dB & 0.436 & 67.40 & 24.41 & 0.372 & 67.53 & 26.23 & \textbf{0.322} & \textbf{69.68} & \textbf{29.17} \\
 & \multirow{3}{*}{Naive-MoDL} & 24dB & 0.197 & \textbf{78.86} & \textbf{32.51} & 0.318 & 61.37 & 28.98 & 1.050 & 24.07 & 18.72 \\
 & & 14dB & 0.228 & 76.65 & 31.53 & 0.303 & 66.89 & 29.32 & 0.638 & 37.75 & 22.47 \\
 & & 4dB & 0.376 & 65.30 & 27.14 & 0.424 & 60.78 & 26.14 & 0.867 & 40.56 & 19.65 \\
\midrule
\end{tabular}%
}
\end{table}
\clearpage

\section{Discussion}
Our study aimed to assess whether denoising of training data could enhance deep learning-based MRI reconstruction quality for noisy, undersampled multi-coil MRI scans across two distinct reconstruction approaches: (1) generative modeling (represented by DPS), and (2) end-to-end supervised learning (represented by MoDL).

Our experiments showed that denoising can enhance reconstruction quality across various SNR levels and anatomies. Notably, in DPS experiments, models trained on GSURE-denoised data at a fixed SNR consistently outperformed Naive-DPS models. An observation from Tables~\ref{tab:brain_metrics} and~\ref{tab:knee_metrics} is that occasionally, GSURE-DPS trained at low SNRs ($12$ dB for brain and $4$ dB for knee), compared to models trained at higher SNRs, often yield better reconstruction performance across all inference SNRs. \add[R1.C8, R1.C11]{Notably, this improvement is not observed in Naive-DPS, whose performance degrades consistently with lower training SNR. This suggests that the benefit is not from training on data corrupted with a certain noise level, but rather from the GSURE-denoising itself. While GSURE-denoising improves effective SNR, models trained on higher-SNR GSURE data (e.g., 32 dB) do not always outperform lower-SNR counterparts, indicating that absolute SNR is not the determining factor. Instead, we hypothesize that GSURE-denoising enables more robust prior learning by correcting signal corruption without over-smoothing, thus exposing the model to realistic yet learnable degradations. This parallels findings from recent works\cite{yaman2020self, aali2024ambient}, where training on corrupted data (appropriately handled) resulted in better generalization under severe inverse problems than training on clean data. We speculate that models trained on clean data may overfit and struggle under distribution shifts, while GSURE-DPS models benefit from exposure to structurally consistent yet imperfect data. While identifying a specific characteristic of GSURE-denoised images that enables robustness remains an open question, this balance between signal preservation and corruption may play a role. Furthermore, since DPS is ultimately an iterative denoising process, the final reconstruction is constrained by the noise level at inference. GSURE-denoised training data may help the model implicitly learn these constraints, enabling more accurate reconstructions even under mismatched noise conditions.}

In addition to improving quantitative metrics, our proposed pipeline demonstrated qualitative robustness, particularly in the knee pathology experiment. GSURE-DPS reconstructions exhibited more graceful degradation at lower inference SNR levels compared to Naive-DPS. This suggests that GSURE-DPS can not only improve accuracy, but also reduce uncertainty in reconstruction outputs.

In the MoDL experiments, we observed that GSURE-MoDL outperformed Naive-MoDL in most scenarios, particularly at low SNR levels. An important observation was that in out-of-distribution cases (low training SNR and high inference SNR), Naive-MoDL performed better. \add[R1.C5, R1.C10]{While the forward model $A$ in MoDL is fixed and known (i.e., it includes the undersampled Fourier transform and coil sensitivities), the data consistency (DC) module and learned regularizer are optimized jointly during training. As a result, MoDL's performance is implicitly coupled to the training noise distribution. This may make it more sensitive to distribution shifts, particularly when inference conditions differ significantly from training. In contrast, GSURE-denoising serves to regularize the training targets, which can reduce noise amplification and improve robustness under such shifts. However, this benefit must be balanced against the risk of introducing signal distortion during denoising.}

Another interesting observation was the difference in performance between brain and knee anatomies at similar SNR levels. For instance, denoising and posterior evaluation metrics for knee scans at $14$ dB consistently displayed worse error metrics compared to brain scans at $12$ dB. We hypothesize that this may stem from our SNR definition based on maximal signal intensity in the image domain, which can vary due to differences in anatomical structures. 

Denoising is a fundamental signal processing step and there is a rich history connecting denoising to image reconstruction\cite{milanfar2024denoising}. At their core, DPMs successively denoise a series of noisy images to arrive at a sample from a target distribution. Through Tweedie's formula\cite{milanfar2024denoising}, there is a direct connection between the MMSE denoiser and the score, which DPMs aim to learn. As a result, it may be possible to consolidate the two-step approach into a single step that directly incorporates the noisy distribution during training\cite{aali2023solving, daras2024consistent}. \add[R2.C3]{However, our results suggest that a two-step approach may provide a more stable and effective training strategy. While a single-step formulation is theoretically appealing, training with both losses simultaneously from scratch may lead to suboptimal convergence, possibly due to increased optimization complexity. Hence, we propose a two-step approach that remains agnostic to both the denoising method and the reconstruction method, allowing flexibility in selecting different denoisers and reconstruction architectures.} 

The learned regularizer in MoDL can also be viewed as a generalized denoising step, where aliasing artifacts are built into the model due to the coupling between the forward model and unrolled iterations. More generally, restoration models may offer a happy medium\cite{hu2024stochastic}. In both cases, the learned denoiser requires a target signal which is assumed to be noise-free. We chose deep neural networks to train the denoisers because of their state-of-the-art performance and the relationship between DPMs and denoisers via Tweedie's rule. \add[R2.C2]{However, other non-DL denoising techniques could also be used, for instance, ones based on wavelet decomposition (SURE-LET\cite{blu2007sure}). We did not explore such approaches, as our goal was to create a training set of noise-free, coil-combined images rather than denoised k-space data. GSURE aligns well with our objective by directly enabling image-space denoising in the presence of correlated noise.}

\begin{figure}
    \centering
    \includegraphics[scale=0.105]{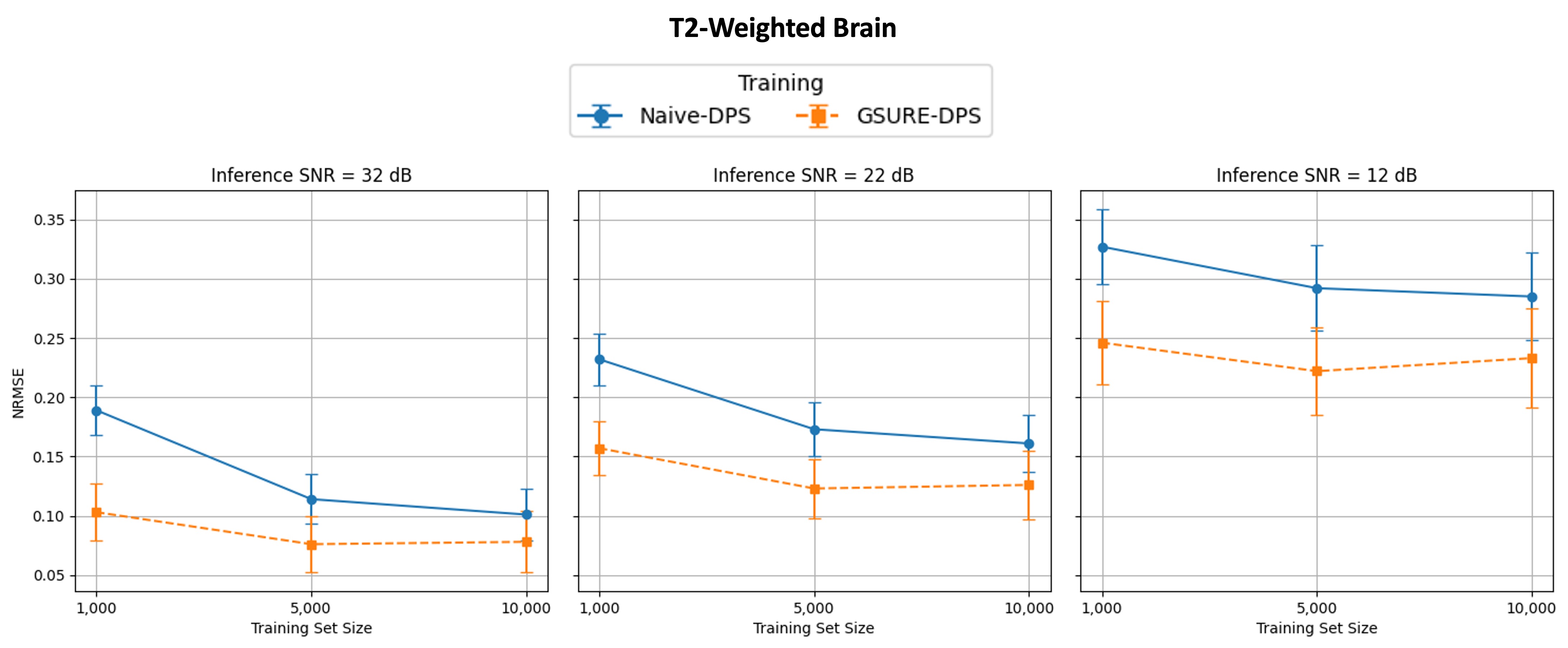}
    \renewcommand{\thefigure}{S6}
    \caption{\scriptsize \add[R1.C4]{Reconstruction performance comparison between Naive-DPS and GSURE-DPS trained on T2-Weighted Brain dataset at 12dB SNR and three different training set sizes. Each plot compares Naive-DPS and GSURE-DPS, where each point on the line represents: a) on y-axis the average NRMSE across 100 validation examples (averaged across 5 random seeds), and b) on x-axis: the training dataset size for the experiment. We can observe that GSURE-DPS reconstructions consistently require less training samples to achieve a target NRMSE, showcasing that GSURE-DPS can improve reconstruction quality and enable faster training.}}
    \label{fig:training_speed}
\end{figure}

\add[R1.C4, R1.C9]{Furthermore, while we specifically chose GSURE for DL-based self-supervised denoising, other self-supervised denoising methods\cite{lehtinen2018noise2noise, batson2019noise2self, tachella2024unsure} also apply the key insight that networks can learn accurate mappings from noisy data, ideally returning an MMSE estimate. Noise2Noise\cite{lehtinen2018noise2noise} uses pairs of noisy images with the same signal but different independent noise realizations. Noise2Self\cite{batson2019noise2self} generalizes this concept beyond the paired image setting using J-invariant denoisers, where the output at a given coordinate does not depend on the input at that coordinate. In this context, GSURE denoising relaxes the J-invariance requirement on the network by regularizing the self-supervised data consistency loss with the network divergence. For J-invariant denoisers, the divergence is equal to zero, reducing the GSURE loss to the data consistency loss\cite{tachella2024unsure}. Furthermore, as discussed in Appendix A.3\cite{lehtinen2018noise2noise}, the expected squared error of the learned target decreases as $\frac{1}{N}$, where $N$ is the number of training samples. This aligns with our empirical finding that GSURE-denoised data enables more data-efficient training by reducing label variance while preserving expectation (Figure~\ref{fig:training_speed}).} Finally, in our work, we assume the available k-space data is always fully sampled. Our work can be extended to handle noisy data that is not fully sampled through ENSURE\cite{aggarwal2022ensure}.

While the benefits of GSURE-based denoising are clear, several limitations require further evaluation. The deep learning-based denoising step introduces additional train-time complexity. Moreover, it is important to recognize the tradeoff between denoising and signal distortion. Denoising can improve signal clarity and reduce noise amplification, ultimately improving reconstruction quality. However, it can also introduce signal distortions, particularly at lower SNRs. Our results indicate that the benefit of denoising outweighs the signal distortion cost when SNR is low. At high SNR, with relatively clean signal, the denoising process may introduce subtle signal distortions. This was evident in certain high-SNR experiments where metrics like SSIM showed slight degradation in GSURE-DPS models compared to Naive-DPS, even as NRMSE and PSNR improved. Balancing denoising strength while maintaining signal integrity remains a challenge for future research. 

Another limitation of our pipeline is that it assumes that the operator $A$ is known. In our work, we derive sensitivity maps using ESPIRiT\cite{ESPIRiT}. This may not be optimal if the target reconstruction uses a different method for calibrating the maps, or if the output is a biased estimate such as RSS reconstruction.

While we used fastMRI to demonstrate a proof of principle, this dataset is already high SNR and does not benefit substantially from denoising (although we show it does not hurt). \add[R2.C4]{The fastMRI dataset does not provide a noise pre-scan. In practice, the noise variance $\sigma_{\eta}$ can be estimated from a pre-scan noise calibration, a common procedure in MRI acquisitions. For our proof-of-concept study, we estimated $\sigma_{\eta}$ from the fully sampled data before retrospective undersampling, though this is not a practical limitation, as real-world implementations can rely on noise measurements from calibration scans.} We envision that our preprocessing will be critical in low-field settings where acquired data (even when fully sampled) are extremely noisy\cite{lyu2023m4raw, shimron2024accelerating}. Recent works have proposed training deep networks in intrinsic SNR units\cite{kellman2005image, xue2024imaging}, providing an approach to mitigate differences across field strengths, SNRs, and coil geometry. Our proposed pipeline could be used as a preprocessing step before training in SNR units.

Our inference speed experiments eluded another limitation of our evaluation approach. Although GSURE-DPS reached target NRMSE with fewer reconstruction steps than Naive-DPS, two images with similar error metrics occasionally exhibited qualitative differences. These metrics are computed against reference images that contain inherent noise, which can obscure true differences in reconstruction quality. This was recently explored in a study\cite{wang2024hidden} which highlighted the "hidden noise" problem in MR reconstruction, suggesting that MRI data contains inherent noise and that conventional error metrics may misalign with the noise characteristics of magnitude MR data. Metrics like NRMSE assume a Gaussian noise distribution, whereas magnitude images follow a noncentral chi distribution, leading to potential misrepresentation of reconstruction quality. Metrics such as the Noncentral Chi Error (NCE) metric proposed by the study\cite{wang2024hidden} better align with the noise distribution in magnitude images, offering a clearer reflection of true image quality. Incorporating such metrics in future work could provide a more reliable evaluation of reconstruction performance, particularly when error metrics are computed on magnitude-data images. These findings also underscore the need for radiologist evaluations to better capture the perceptual quality of the reconstructions.

\section{Conclusion}
\label{sec:conclusion}
This study demonstrated the effect of incorporating GSURE-based self-supervised denoising as a pre-processing step to improve deep learning-based MRI reconstruction, particularly in challenging low-SNR scenarios. By improving the quality of training data, our proposed pipeline facilitates the learning of more robust and efficient deep learning-based reconstruction methods. Although the benefits of denoising are pronounced, especially in low-SNR environments, our findings highlight the need for further optimization to address computational complexity. Furthermore, while quantitative improvements in uncertainty reduction and reconstructions suggest that GSURE denoising has the potential to enhance MRI reconstruction, these outcomes require qualitative verification from expert radiologists. Overall, our results provide a strong foundation for future research and clinical translation, offering a pathway toward more accurate and efficient MRI reconstructions.

\section{Acknowledgments}
This work was supported by NSF IFML 2019844, NSF CCF-2239687
(CAREER), NIH U24EB029240, Google Research Scholar Program, and Oracle for Research Fellowship. J.T. would like to thank Dr. Frank Ong for first introducing him to SURE.

\section{Data and Code Availability Statement}
We open-source and release our source code: \href{https://github.com/utcsilab/gsure-diffusion-mri.git}{github.com/utcsilab/gsure-diffusion-mri}.

\section{Conflicts of Interest}
The authors do not have any conflicts of interest.

\section{ORCID}
\textit{Asad Aali} \url{https://orcid.org/0009-0008-2120-5722}\\
\noindent \textit{Marius Arvinte} \url{https://orcid.org/0000-0002-9328-9223}\\
\noindent\textit{Jonathan I.\ Tamir} \url{https://orcid.org/0000-0001-9113-9566}

\clearpage
\printbibliography

\clearpage

\begin{table}[htbp]
\centering
\renewcommand{\thetable}{S1}
\caption{\scriptsize Validation performance of GSURE denoising averaged across 100 examples, for three SNR levels and two anatomies: a) T2-Weighted Brain and b) Fat-Suppressed Knee.}
\label{tab:brain_knee_denoising_metrics}
\vspace{0.1in}
\begin{tabular}{c|c|c|c|c}
\toprule
\textbf{Anatomy} & \textbf{Dataset SNR} & \textbf{NRMSE $\downarrow$} & \textbf{SSIM $\uparrow$} & \textbf{PSNR $\uparrow$} \\
\midrule
\multirow{3}{*}{Brain} & 32dB & 0.065 $\pm$ 0.027 & 97.39 $\pm$ 0.77 & 38.86 $\pm$ 4.13 \\
& 22dB & 0.089 $\pm$ 0.025 & 94.61 $\pm$ 0.96 & 36.59 $\pm$ 2.67 \\
& 12dB & 0.115 $\pm$ 0.030 & 91.89 $\pm$ 1.30 & 34.73 $\pm$ 1.50 \\
\midrule
\multirow{3}{*}{Knee} & 24dB & 0.107 $\pm$ 0.027 & 91.80 $\pm$ 3.61 & 37.51 $\pm$ 2.41 \\
& 14dB & 0.179 $\pm$ 0.042 & 77.64 $\pm$ 7.93 & 34.20 $\pm$ 2.60 \\
& 4dB  & 0.309 $\pm$ 0.095 & 73.25 $\pm$ 8.38 & 30.54 $\pm$ 2.60 \\
\bottomrule
\end{tabular}%
\end{table}

\clearpage

\begin{figure}
    \centering
    \includegraphics[scale=0.095]{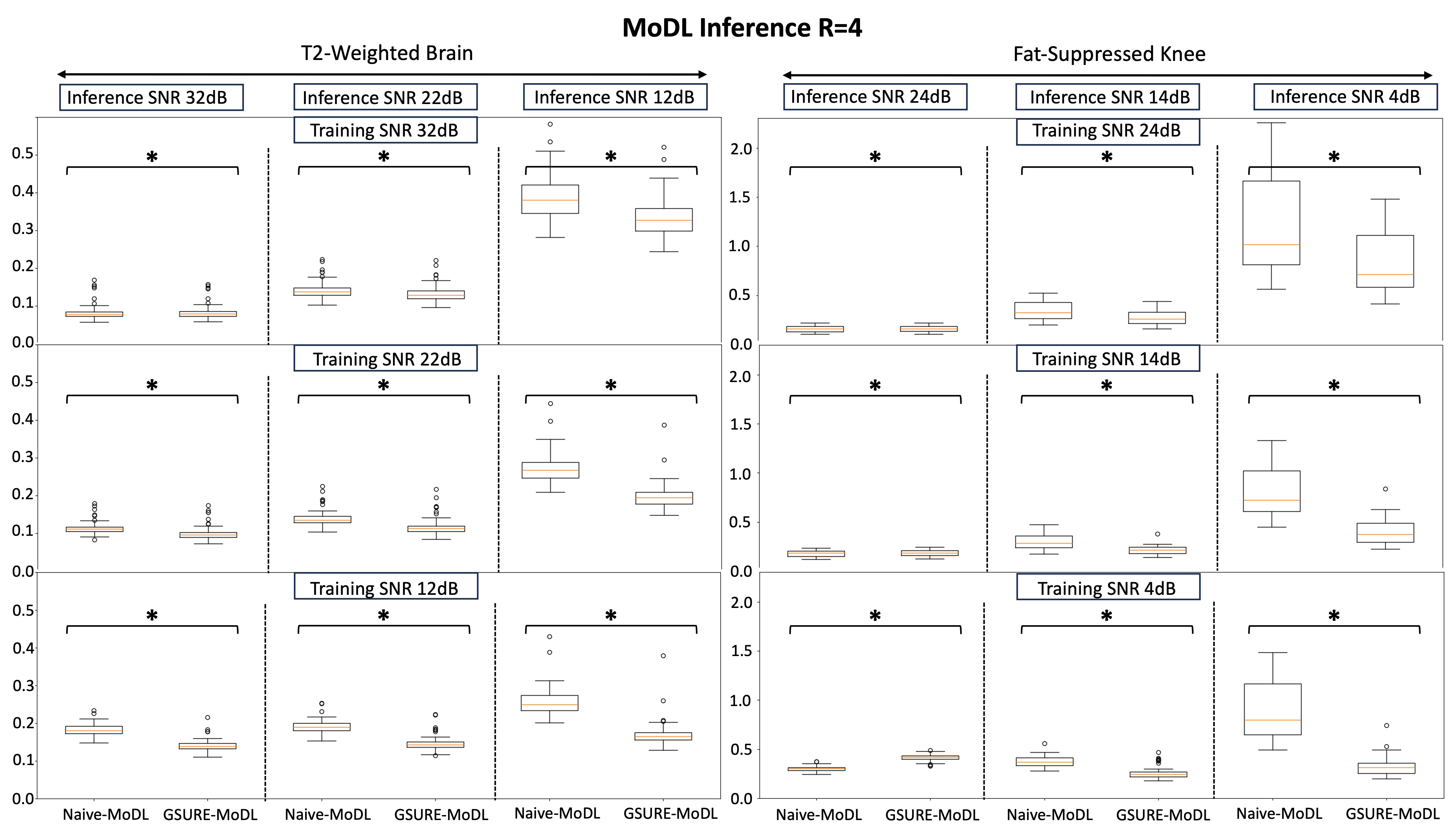}
    \renewcommand{\thefigure}{S4}
    \caption{\scriptsize Box plot of $100$ validation reconstruction examples (NRMSE) comparing Naive-MoDL and GSURE-MoDL at acceleration factor $R=4$, across: a) T2-Weighted Brain and Fat-Suppressed anatomies, b) training SNR levels, c) inference SNR levels. (\textbf{$^*$}statistically significant difference in reconstruction performance, NS: not significant).
}
    \label{fig:modl_r4}
\end{figure}

\begin{figure}
    \centering
    \includegraphics[scale=0.095]{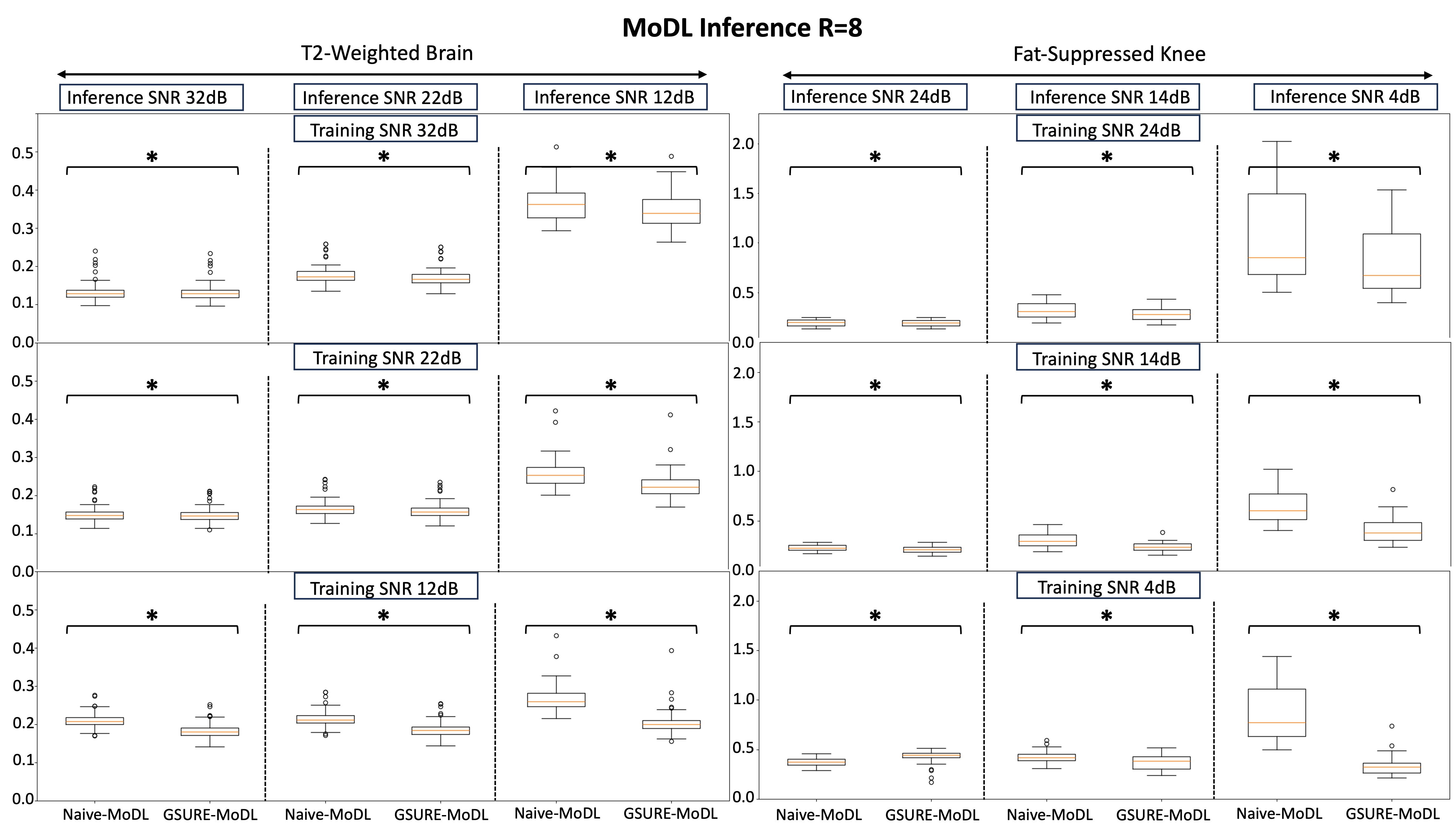}
    \renewcommand{\thefigure}{S5}
    \caption{\scriptsize Box plot of $100$ validation reconstruction examples (NRMSE) comparing Naive-MoDL and GSURE-MoDL at acceleration factor $R=8$, across: a) T2-Weighted Brain and Fat-Suppressed anatomies, b) training SNR levels, c) inference SNR levels. (\textbf{$^*$}statistically significant difference in reconstruction performance, NS: not significant).
}
    \label{fig:modl_r8}
\end{figure}

\clearpage

\clearpage

\clearpage

\begin{figure}
    \centering
    \includegraphics[scale=0.26]{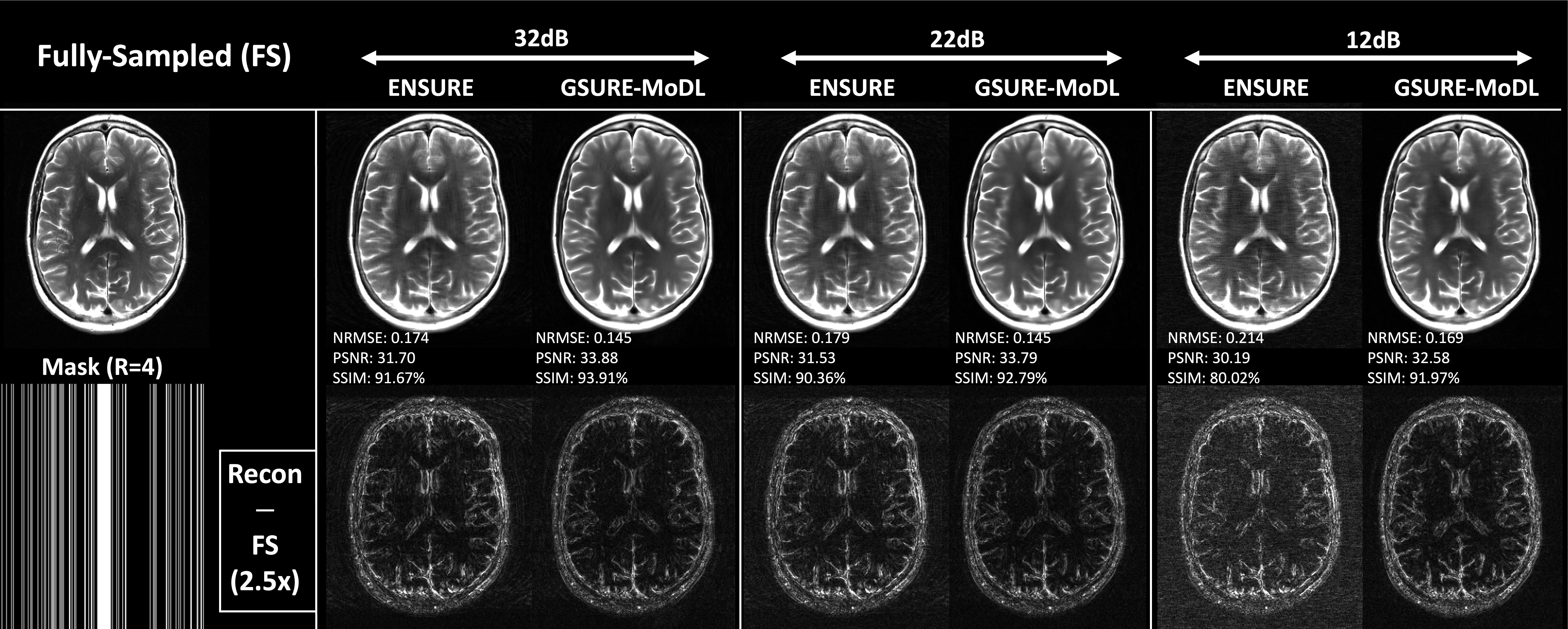}
    \renewcommand{\thefigure}{S9}
    \caption{\scriptsize \add[R2.C3]{T2-Weighted Brain reconstructions with ENSURE and GSURE-MoDL models trained on $12$dB SNR data. Across columns, we show reconstructions across three inference SNR levels. In the first row, we show the reconstruction example with quantitative comparison metrics. In the second row, we show the difference between the reconstruction and fully sampled image at $2.5\times$ brightness. We can observe that GSURE-MoDL outperforms ENSURE, notably at lower SNR levels.}}
    \label{fig:brain_ensure}
\end{figure}

\begin{table}[htbp]
\centering
\renewcommand{\thetable}{S2}
\caption{\scriptsize \add[R2.C3]{T2-Weighted Brain reconstruction metrics in comparison to ENSURE. Across each inference SNR, acceleration factor, and reconstruction method, we \textbf{highlight} the training strategy with the best reconstruction performance (lowest error).}}
\label{tab:ensure_metrics}
\vspace{0.1in}
\resizebox{\textwidth}{!}{%
\begin{tabular}{ccc|ccc|ccc|ccc}
\toprule
\multirow{3}{*}{\shortstack{\textbf{Acceleration}\\\textbf{Factor}}} & \multirow{3}{*}{\shortstack{\textbf{Reconstruction}\\\textbf{Method}}} & \multirow{3}{*}{\shortstack{\textbf{Training}\\\textbf{SNR}}} & \multicolumn{9}{c}{\shortstack{\textbf{Inference SNR}}} \\
\cmidrule(lr){4-12}
 & & & \multicolumn{3}{c}{\textbf{32dB}} & \multicolumn{3}{c}{\textbf{22dB}} & \multicolumn{3}{c}{\textbf{12dB}} \\
\cmidrule(lr){4-6} \cmidrule(lr){7-9} \cmidrule(lr){10-12}
& & & \textbf{NRMSE $\downarrow$} & \textbf{SSIM $\uparrow$} & \textbf{PSNR $\uparrow$} & \textbf{NRMSE $\downarrow$} & \textbf{SSIM $\uparrow$} & \textbf{PSNR $\uparrow$} & \textbf{NRMSE $\downarrow$} & \textbf{SSIM $\uparrow$} & \textbf{PSNR $\uparrow$} \\
\midrule
\multirow{3}{*}{4} & Naive-MoDL & 12dB & 0.183 & 81.89 & 30.99 & 0.192 & 80.32 & 30.56 & 0.256 & 70.58 & 27.69 \\
 & ENSURE & 12dB & 0.167 & 89.64 & 30.60 & 0.174 & 88.71 & 30.44 & 0.208 & 79.66 & 29.25 \\
 & GSURE-MoDL & 12dB & \textbf{0.141} & \textbf{90.85} & \textbf{32.58} & \textbf{0.146} & \textbf{90.33} & \textbf{32.46} & \textbf{0.169} & \textbf{89.50} & \textbf{31.48} \\
\midrule
\multirow{3}{*}{8} & Naive-MoDL & 12dB & 0.211 & 78.82 & 29.03 & 0.215 & 78.04 & 28.94 & 0.265 & 70.83 & 27.11 \\ 
 & ENSURE & 12dB & 0.216 & 84.85 & 28.17 & 0.219 & 84.55 & 28.12 & 0.238 & 80.36 & 27.68 \\
 & GSURE-MoDL & 12dB & \textbf{0.183} & \textbf{86.19} & \textbf{29.84} & \textbf{0.186} & \textbf{86.47} & \textbf{29.79} & \textbf{0.204} & \textbf{86.76} & \textbf{29.29} \\
\midrule
\end{tabular}%
}
\end{table}

\end{document}